\documentclass[aps,prd,twocolumn,superscriptaddress,floatfix]{revtex4-1}

\usepackage{amsmath}
\usepackage{bm}
\usepackage{mathptmx}
\usepackage{color}
\usepackage{helvet}
\usepackage{courier}
\usepackage{mathrsfs}
\usepackage[dvips]{graphicx}
\usepackage{booktabs}
\usepackage{ulem}

\DeclareMathAlphabet{\mathcal}{OMS}{cmsy}{m}{n}

\begin{document}


\title{Flavor structure of $\Lambda$ baryons from lattice QCD: From strange to charm quarks}



\author{Philipp Gubler}
\email[]{pgubler@riken.jp}
\affiliation{Institute of Physics and Applied Physics, Yonsei University, Seoul 120-749, Korea}

\author{Toru T. Takahashi}
\affiliation{Gumma National College of Technology, Gumma 371-8530, Japan}

\author{Makoto Oka}
\affiliation{Tokyo Institute of Technology, Department of Physics, Meguro, Tokyo 152-8551, Japan} 
\affiliation{Advanced Science Research Center, JAEA, Tokai, Ibaraki, 319-1195, Japan} 


\date{\today}

\begin{abstract}
We study $\Lambda$ baryons of spin-parity $\frac{1}{2}^{\pm}$ with either a strange or charm valence quark 
in full 2+1 flavor lattice QCD. Multiple $SU(3)$ singlet and octet operators are employed to generate the desired single 
baryon states on the lattice. Via the variational method, the couplings of these states to the different operators provide information 
about the flavor structure of the $\Lambda$ baryons. 
We make use of the gauge configurations of the PACS-CS Collaboration and chirally extrapolate the results for the masses and $SU(3)$ 
flavor components to the physical point. 
We furthermore gradually change the hopping parameter of the heaviest quark from strange to charm to study how the properties of the 
$\Lambda$ baryons evolve as a function of the heavy quark mass. It is found that the baryon energy levels increase almost linearly with the 
quark mass. Meanwhile, the flavor structure of most of the states remains stable, with the exception of the lowest $\frac{1}{2}^{-}$ state, 
which changes from a flavor singlet $\Lambda$ to a $\Lambda_c$ state with singlet and octet components of comparable size. 
Finally, we discuss whether our findings can be interpreted with the help of a simple quark model and find that the negative-parity $\Lambda_c$ states  
can be naturally explained as diquark excitations of the light $u$ and $d$ quarks. On the other hand, the quark-model picture does not appear to 
be adequate for the negative-parity $\Lambda$ states, suggesting the importance of other degrees of freedom to describe them. 
\end{abstract}

\pacs{}

\maketitle

\section{Introduction}

The lightest $J^P=1/2^-$ $\Lambda$ baryon, 
$\Lambda(1405)$, has been of great interest from several points of view. 
In spite of its valence strange quark, 
$\Lambda (1405)$ is the lightest among the negative-parity baryons, and is 
especially much lighter than
its nonstrange counterpart $N(1535)$.
The structure of $\Lambda (1405)$ is also under dispute.
While it is interpreted as a flavor-singlet state
in terms of the flavor $SU(3)$ symmetry, 
the $\Lambda (1405)$ could be regarded as a $\overline{K}N$ molecular bound state,
which would require no spin-orbit partner. 
In this case, the bound state's 
binding energy of $\sim 30$ MeV implies a
strong attraction between $\overline{K}$ and $N$
~\cite{Sakurai:1960ju,Dalitz:1967fp}, 
which has led to predictions of kaonic nuclei or kaonic nuclear matter
~\cite{Akaishi:2002bg,Yamazaki:2002uh,Akaishi:2005sn}. 
The $\Lambda (1405)$ has furthermore been conjectured to consist of two poles, 
which are respectively dominated by $\overline{K}N$ and $\pi\Sigma$ components~\cite{Jido:2003cb,Hyodo:2007jq,Ikeda:2012au}.

Lattice QCD is a powerful nonperturbative tool,
which enables us to clarify the strong interactions
in a model-independent way  based on QCD.
Several lattice QCD studies
on $\Lambda (1405)$ have been performed so far
~\cite{Melnitchouk:2002eg,Nemoto:2003ft,Burch:2006cc,Ishii:2007ym,Takahashi:2009bu,Menadue:2011pd,Engel,Hall},
and 
the signal of $\Lambda$(1405) was recently identified~\cite{Menadue:2011pd}.
In a subsequent paper~\cite{Hall},
the electromagnetic response of the $\Lambda$ was investigated, 
and it was conjectured that
the strange quark in $\Lambda$(1405) is confined in a spin-0 state, that is, the kaon. 
This evidence for the $\overline{K}N$-molecular picture of the $\Lambda$(1405) is of interest, as it may account 
for its mysterious properties. The key concept here is the {\it flavor symmetry}.

Then, how does the flavor-based property emerge? 
One may recall the $\Lambda_c$ baryons,
which are the counterparts of $\Lambda$
that contain the much heavier charm quark.
The flavor symmetry is therefore largely broken,
and its nature should be quite different from $\Lambda$ 
(for recent lattice studies about charmed baryons and their flavor structure, see Ref.~\cite{Bali:2015lka} and the 
references cited therein). 
The key symmetry here would be the {\it heavy quark symmetry},
which reflects the fact that
spin-spin interactions are suppressed 
between light and heavy quarks.
The connection between $\Lambda$ and $\Lambda_c$ was recently investigated
using a simple quark model~\cite{Yoshida},
and it was found that in the $\Lambda_c$ baryons
the diquark degrees of freedom emerge and that their low-lying spectrum 
can be naturally explained in terms of diquarks.

In this paper, we study the properties of  $\Lambda$ baryons
with 2+1 flavor lattice QCD, adopting the flavor $SU(3)$ ``octet'' and ``singlet''
baryonic operators,
which enables us to clarify 
the flavor structure of the $\Lambda$ and $\Lambda_c$ baryons.
By gradually evolving the strange into the charm quark mass, 
we interpolate between $\Lambda$ and $\Lambda_c$,
and hence systematically investigate the structural change of the $\Lambda$ particles. 

The paper is organized as follows. In Sec. \ref{Setup}, we briefly explain our 
lattice setup, the employed interpolating fields and the variational method used to extract the eigenenergies 
of the states as well as their flavor content. In Sec. \ref{Results}, the obtained $\Lambda$ baryon 
spectrum and the respective flavor decomposition are presented, while in Sec. \ref{Discussion}, we discuss 
how these results can (or cannot) be interpreted in a quark-model context. A summary and 
conclusions follow in Sec. \ref{SummaryConclusion}. Finally, the numerical results are summarized in Appendix A. 

\section{\label{Setup} Lattice QCD setup}

\subsection{Simulation conditions}

We adopt the renormalization-group-improved action for gauge fields
and the ${\mathcal O}(a)$-improved action for quarks.
The coupling $\beta$ in the gauge action is $\beta = 1.9$, 
the corresponding lattice spacing is $a = 0.0907$ fm~\cite{Aoki:2008sm}, 
and the lattice size is $32^3 \times 64$. 
The hopping parameters 
for the strange quark $\kappa_s$ and the charm quark $\kappa_c$
are set to be 0.13640 and 0.1224,
and those for light quarks $\kappa$ are 
0.13700, 0.13727, 0.13754, and 0.13770, with 
the corresponding pion masses ranging approximately from 700 MeV to 290 MeV. 

\subsection{Baryonic operators for spin 1/2}

The low-lying $\Lambda$ states in the $S=-1$ and $I=0$ channel
are extracted from $4\times 4$ cross-correlators.
For generating the $\Lambda$ states, we adopt the following isosinglet operators: 
\begin{eqnarray}
\Lambda_{\mu_1\mu_2\mu_3} 
=
\frac{\varepsilon_{abc}}{\sqrt 2}
\left(
u_{\mu_1}^a d_{\mu_2}^b
-
d_{\mu_1}^a u_{\mu_2}^b
\right)
Q_{\mu_3}^c.
\end{eqnarray}
In the case of $\Lambda$ ($\Lambda_c$) baryons, $Q_{\mu_3}^c$
is the strange-quark field $s_{\mu_3}^c$ (the charm-quark field $c_{\mu_3}^c$).
The spinor indices $\mu$ are taken according to the classification
in Table VIII in Ref.\,\cite{Basak:2005ir}, 
where $\overline{\Psi}^{G_{1g/u,i}}_{1/2} (i=1,2,3)$ are flavor-octet operators
and 
$\overline{\Psi}^{G_{1g/u,4}}_{1/2}$ flavor-singlet operators.
''Flavor-singlet (octet) operators'' here means that 
they belong to the flavor-singlet (octet) irreducible representation of the $SU(3)_f$
symmetry when all the quark masses are equal ($m_u=m_d=m_Q$).
Note that we always consider $SU(3)_f$ flavor symmetry for three quark fields, $u$, $d$ and $Q$.
We eventually have three octet operators and one singlet operator 
for spin-1/2 $\Lambda$ states.

\subsection{Flavor content and eigenenergies of $\Lambda$ states}

One important goal of this paper is the clarification
of the flavor content in low-lying $\Lambda$ states,
which can be extracted via the diagonalization of cross-correlators. 
Let us consider a situation where we have a set of $N$ independent operators.
We define cross-correlators as
\begin{equation}
{\mathcal M}(x,y)_{IJ}
\equiv
\langle
{\eta}_I(x)
\overline{\eta}_J(y)
\rangle,
\end{equation}
for positive- and negative-parity channels,
where the operators $\eta_I$ denote
quasilocal spin-1/2 operators of positive or negative parity, 
\begin{eqnarray}
\overline{\eta}_I
\equiv
\overline{\Psi}^{G_{1g/u,I}}_{1/2}.
\end{eqnarray}
The subscript $I$ denotes the operator type
in terms of the irreducible representation of the octahedral group.
(Flavor-octet for $I=1,2,3$ and flavor-singlet for $I=4$.)
We adopt gauge-invariant smeared operators for sources and sinks.
Smearing parameters are chosen so that the 
root-mean-square radius is approximately 1.0 fm. 

Then, correlation matrices
${\mathcal M}_{IJ}(t)\equiv\langle \eta_I(t)
\overline{\eta}_J(0)\rangle$ can be
decomposed into the sum over the energy eigenstates $|i \rangle$ as
\begin{eqnarray}
{\mathcal M}_{IJ}(t)
&\equiv&
{\mathcal M}_{IJ}(t,0)
=
\langle \eta_I(t) \overline{\eta}_J(0)\rangle
\nonumber \\
&=&
\sum_{i,j}
(C^\dagger_{{\rm snk}})_{Ii}
\Lambda(t)_{ij}
(C_{{\rm src}})_{jJ} \nonumber \\
&=&
(C^\dagger_{\rm snk} \Lambda(t) C_{\rm src})_{IJ},
\end{eqnarray}
where the lowercase letters ($ij$)
are the indices for the intermediate
energy eigenstates.
Here, the diagonal matrix $\Lambda(t)$ is defined as
\begin{equation}
\Lambda(t)_{ij}\equiv \delta_{ij} e^{- E_it},
\label{dmatrix}
\end{equation}
and the coefficients
\begin{eqnarray}
(C^\dagger_{\rm snk})_{Ii}\equiv \langle {\rm vac} | \eta_I | i \rangle, 
\quad
(C_{\rm src})_{jI}\equiv \langle j | \overline{\eta}_J | {\rm vac} \rangle, 
\end{eqnarray}
are the couplings between $\Lambda$ operators and energy eigenstates.
We define the $I$th operator's overlap $\psi_{Ii}$ to the $i$th spin-1/2 $\Lambda$ state
by the coupling $\langle {\rm vac} | \eta_I | i \rangle$, 
\begin{eqnarray}
\psi_{Ii}^{\frac12}
\equiv
\langle {\rm vac} | \eta_I | i \rangle
=
(C^\dagger_{\rm snk})_{Ii}, 
\label{eq:overlap}
\end{eqnarray}
which in this paper is used to measure the flavor content of each $\Lambda$ state.

The eigenenergy of each state $E_i$ and their corresponding couplings
$\psi_{Ii}^{\frac12}$ can be extracted by diagonalizing the correlation matrix.
From the product 
\begin{equation}
{\mathcal M}^{-1}(t+1){\mathcal M}(t)
=C_{\rm src}^{-1}\Lambda(-1)C_{\rm src},
\end{equation}
one can extract the eigenenergies $E_i$
from the eigenvalues $e^{E_i}$ 
of ${\mathcal M}^{-1}(t+1){\mathcal M}(t)$. 

Modulo overall constants,
$(C_{\rm src})^{-1}$ and $(C^\dagger_{\rm snk})^{-1}$
can be obtained as right and left eigenvectors
of 
${\mathcal M}^{-1}(t+1){\mathcal M}(t)$ and ${\mathcal M}(t){\mathcal M}(t+1)^{-1}$, 
respectively,
since
\begin{equation}
{\mathcal M}^{-1}(t+1){\mathcal M}(t)(C_{\rm src})^{-1}
=(C_{\rm src})^{-1}\Lambda(-1)
\end{equation}
and
\begin{equation}
(C^\dagger_{\rm snk})^{-1}{\mathcal M}(t){\mathcal M}(t+1)^{-1}
=\Lambda(-1)(C^\dagger_{\rm snk})^{-1}
\end{equation}
hold.

In the actual calculation of the eigenenergies, 
to avoid unstable diagonalization at large $t$, 
we determine the couplings at relatively small $t$ and construct optimal source and sink operators,
${\cal O}^{\rm src \dagger}_i(t)$ and 
${\cal O}^{\rm snk}_i(t)$,
which couple dominantly (solely in the ideal case) to the $i$th lowest state,
as
\begin{equation}
{\cal O}^{\rm src \dagger}_i(t)=\sum_J
\overline{\eta}_J(t) (C_{\rm src})^{-1}_{Ji}
\end{equation}
and
\begin{equation}
{\cal O}^{\rm snk}_i(t)=\sum_J
(C^\dagger_{\rm snk})^{-1}_{iJ} \eta_J(t).
\end{equation}
In fact, their correlation function leads to a single-exponential form, 
\begin{equation}
\langle
{\cal O}^{\rm snk}_i(t)
{\cal O}^{\rm src \dagger}_i(0) 
\rangle
=e^{-E_i t}, 
\label{diag01}
\end{equation}
where we have ignored contributions of states that are above the lowest $N$ eigenstates. 
We note here that,
if the correlation matrix ${\mathcal M}(t)$ is Hermitian,
one can determine
$(C_{\rm src})^{-1}$ and $(C_{\rm snk})^{-1}$ up to overall phase factors
so that Eq.(\ref{diag01}) is satisfied.


\section{\label{Results} Lattice QCD Results}
\subsection{Hadron masses}
\begin{figure*}
\begin{center}
\centering
\includegraphics[width=7.8cm]{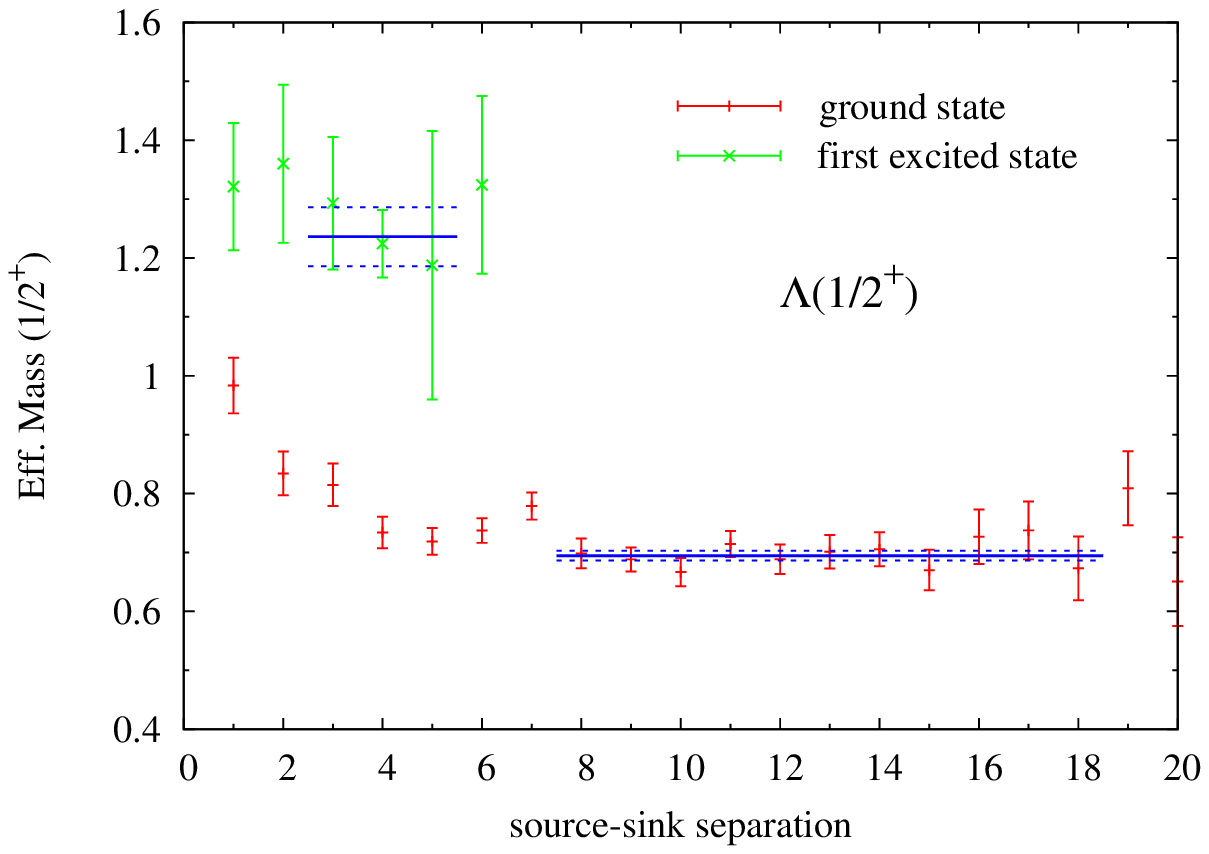}
\includegraphics[width=7.8cm]{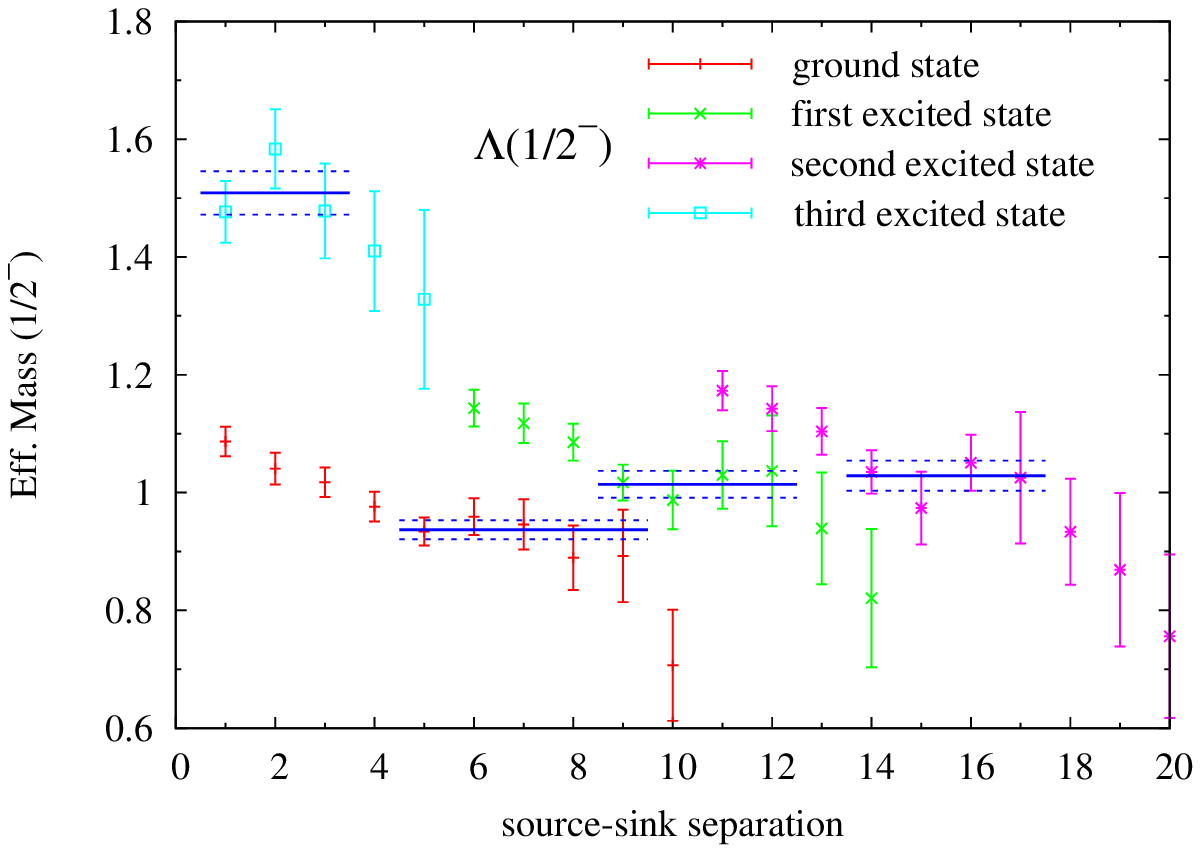}
\\
\includegraphics[width=7.8cm]{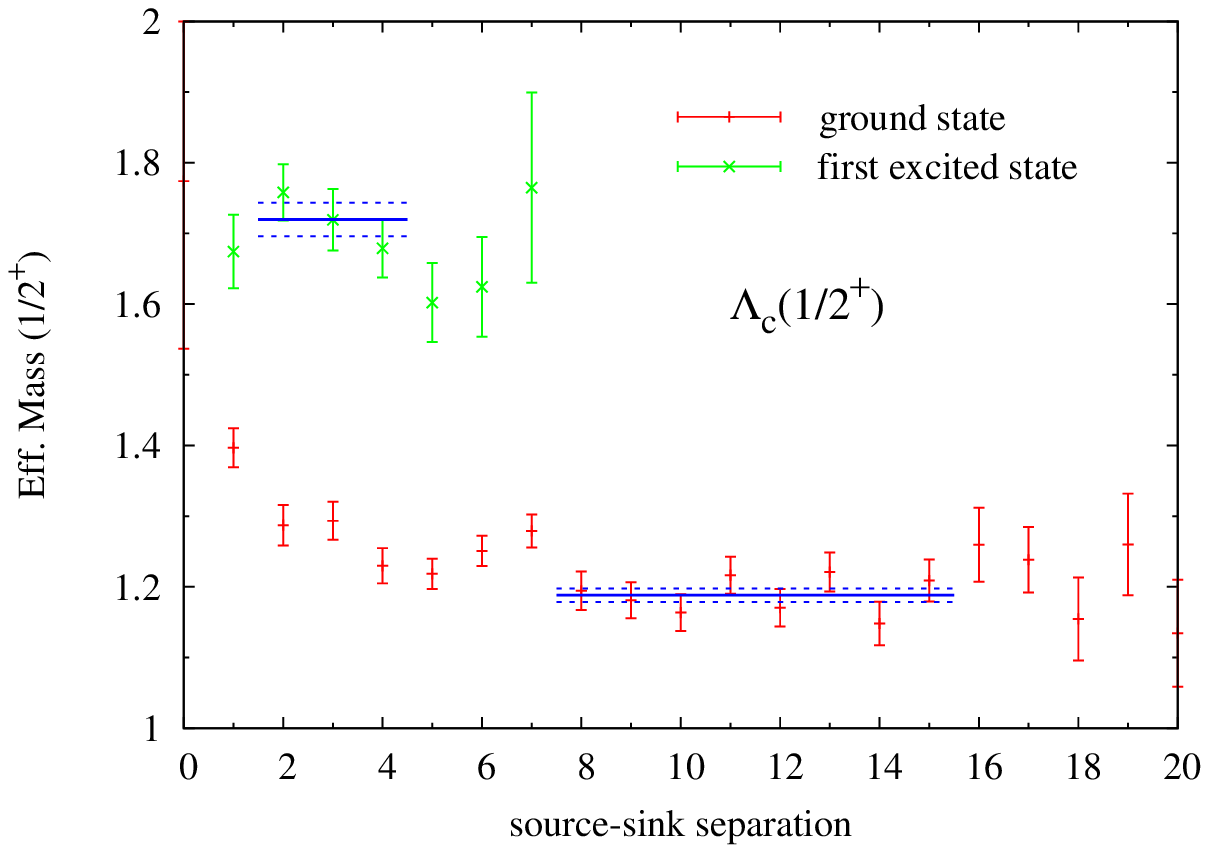}
\includegraphics[width=7.8cm]{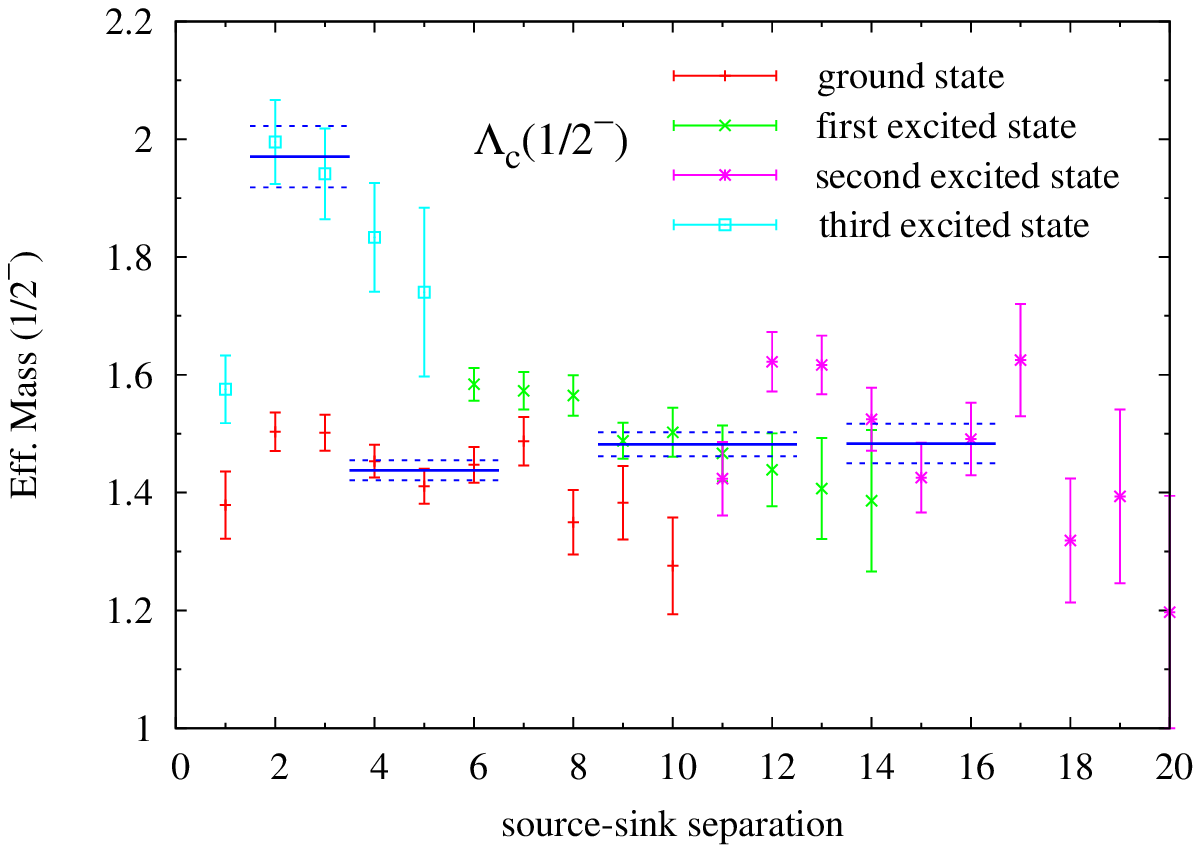}
\caption{Effective mass plots for the positive-parity (left plots) and negative-parity (right plots) channels for both $\Lambda$ (upper plots) and $\Lambda_c$ (lower plots). 
The positive-parity plots show the ground and first-excited states, while for the negative-parity case, ground, first-excited, second-excited, and third-excited states 
are shown. To improve the visibility of the negative-parity plots, we have horizontally shifted the data points of the first- (second-) excited state 
by 5 (10) time slices. 
The horizontal dark blue solid and dashed lines indicate the fit results within the plateau region for each channel. 
All plots are given in lattice units.}
\label{fig:eff.mass}
\end{center}
\end{figure*}

Let us first show a few representative effective mass plots in Fig.\ref{fig:eff.mass}. 
They are given for $\kappa_{ud} = 0.13727$, with $\kappa_s = 0.13640$ for $\Lambda$ and $\kappa_c = 0.12240$ for $\Lambda_c$. 
The eigenenergies of each state of each channel are obtained by a fit to the data points in the respective plateau regions. 
The errors shown here are purely statistical and are obtained from a singly binned jackknife analysis of the lattice data. 
The fitting results are 
indicated as dark blue solid and dashed lines in Fig.\ref{fig:eff.mass}. 
We have checked that increasing or decreasing the upper boundary of the plateau 
region by a few time slices does not alter the effective mass averages beyond their statistical 
errors. 
While increasing the lower boundary similarly always gives statistically
consistent masses, decreasing it by more than one time slice
typically leads to a statistically significant increase of the mass
average, which indicates that the lower boundaries adopted in this paper
are the smallest that are safe from contaminations of the high excited states. 
We have studied altogether four different $\kappa_{ud}$ values ($0.13700$, $0.13727$, $0.13754$, and $0.13770$) and 
five hopping parameters for the heavy quark ($\kappa_s$ and $\kappa_c$ given above and three values that interpolate 
between strange and charm: $\kappa_{sc} = 0.13300$, $0.12900$, and $0.12600$). 
To extrapolate the results to the physical point, we have performed a quadratic fit 
to all four available data points. 
To get an idea about the systematic uncertainties of this procedure, we have also carried out a linear fit 
to the data, for which the results are given in Appendix A. It is seen there that in some cases the difference between 
linear and quadratic extrapolation results exceeds the statistical error, which shows that the systematic uncertainty of the 
chiral extrapolation is not negligible. 
To determine the two-particle thresholds under various conditions, we have extracted the energies of 
the mesonic $\pi$, $\overline{K}$, $D$, and baryonic $N$, $\Sigma$, $\Sigma_c$ states. 
The numerical results are summarized in Tables \ref{tab:num.eff.masses.1},  \ref{tab:num.eff.masses.2}, and \ref{tab:num.eff.masses.3} of Appendix A.

We will now look at the $\Lambda$ baryons in detail. 
It is clearly seen 
already in the plots of Fig. \ref{fig:eff.mass} that the level structures of both $\Lambda$ and $\Lambda_c$ exhibit the same qualitative behavior, which 
is simply shifted due to the large charm quark mass. This similarity is seen for all hopping parameters $\kappa_{sc}$ that we have investigated in this work. 
For positive parity, there is a large gap between the ground and the first excited state, while for negative parity, we find the first two excited states close to 
the ground state and the third excited state about 500 MeV above the lowest three states. 

\begin{figure*}
\begin{center}
\centering
\includegraphics[width=7.8cm]{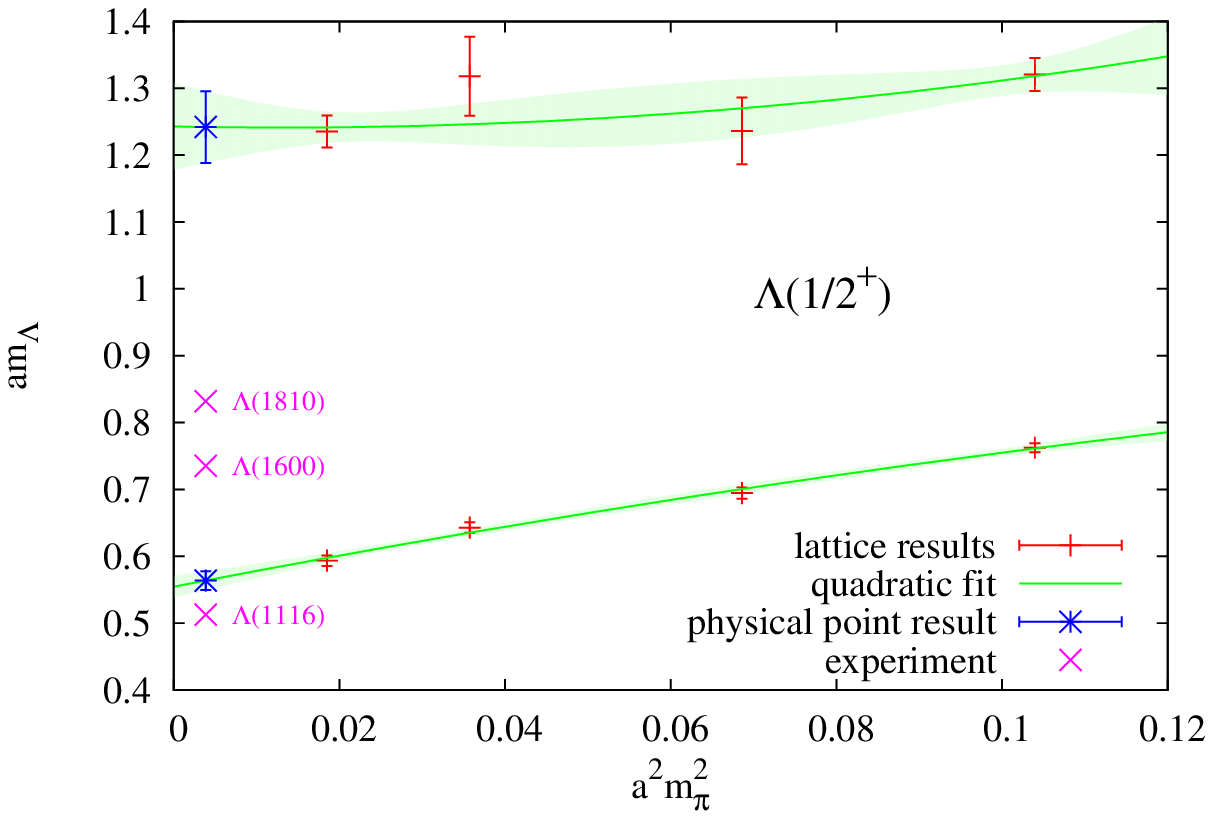}
\includegraphics[width=7.8cm]{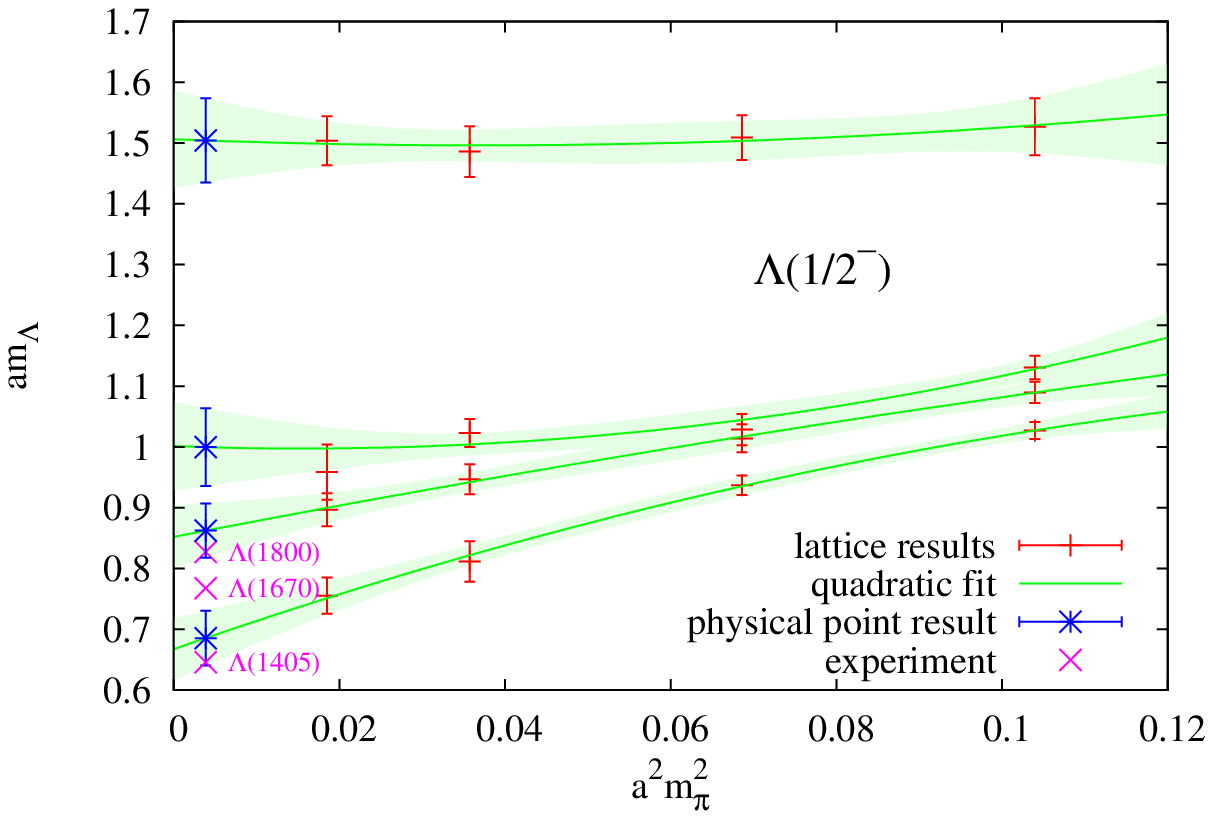}
\\
\includegraphics[width=7.8cm]{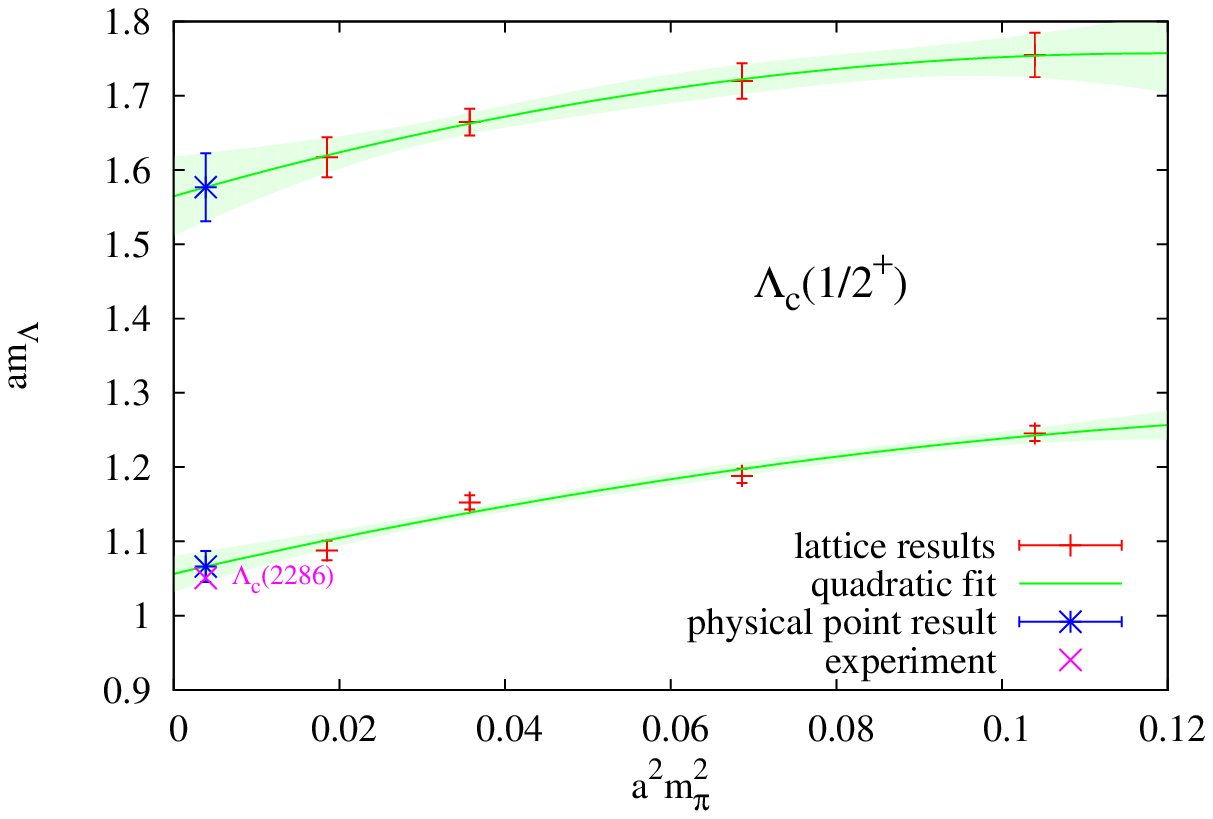}
\includegraphics[width=7.8cm]{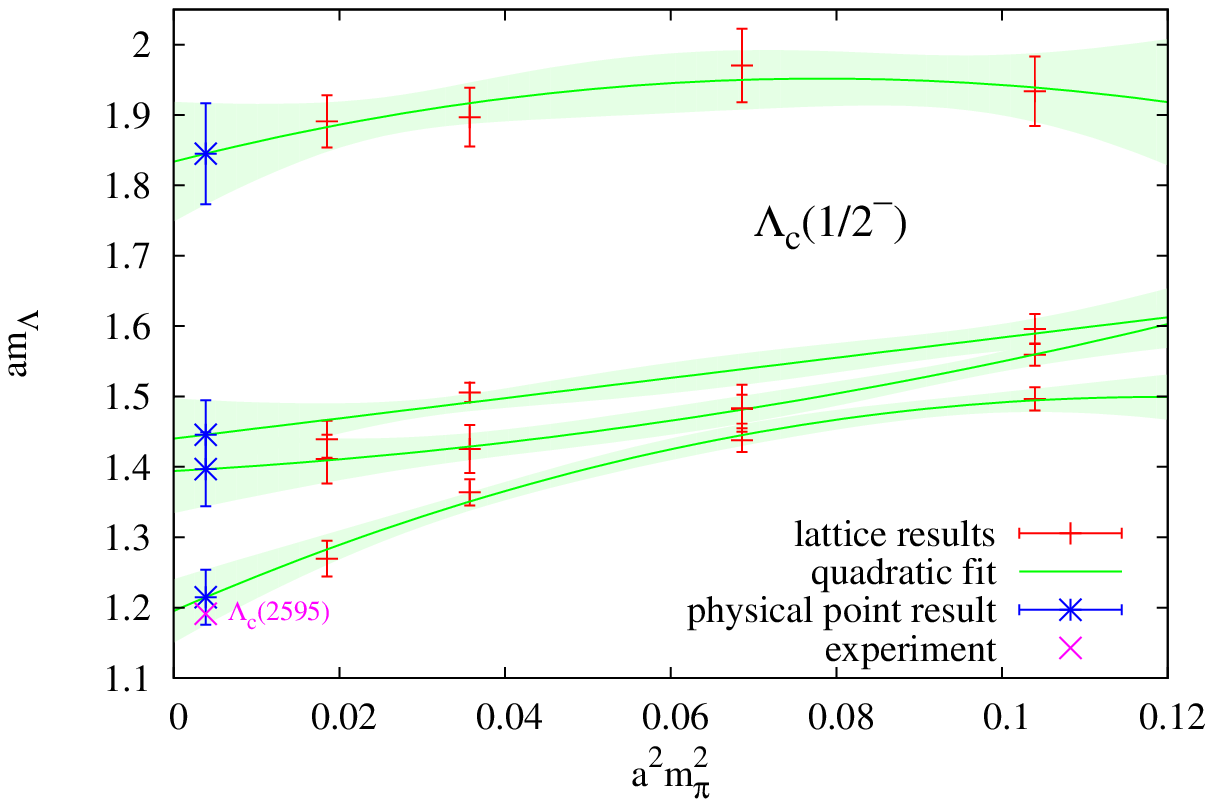}
\caption{Masses of the $\Lambda$ baryons of positive (left plots) and negative parity (right plots) containing a strange (upper plots) or charm quark (lower plots), 
shown as a function of the squared pion mass. The green curves are quadratic fits to the lattice points (shown in red). The blue points give the extrapolated physical point masses and 
the pink points the experimentally observed $\Lambda$ baryon spectrum. 
All plots are given in lattice units.}
\label{fig:extrapolation}
\end{center}
\end{figure*}

To see how the energy levels behave as a function of the squared pion mass (or, equivalently, the $u$- and $d$-quark masses), we plot 
the eigenenergies in Fig. \ref{fig:extrapolation} as a function of $a^2 m^2_{\pi}$. Here, we again only show the cases corresponding to the ``physical''
hopping parameters $\kappa_s$ and $\kappa_c$. For the values interpolating between these two, a similar behavior is observed. 
In Fig. \ref{fig:extrapolation}, we furthermore show 
the quadratic extrapolations as green lines and the respective errors as green shaded areas. 
The values of the experimental hadron masses for each channel are plotted as pink crosses, 
which should be compared to our extrapolated physical point results (shown in blue). 

For the $\Lambda$ states with a strange quark, we observe that the ground states of both positive and negative parity are extracted close to, but consistently above, the 
experimental values. A similar result was obtained in Ref.~\cite{Menadue:2011pd}, in which only the negative-parity $\Lambda$ baryons were studied and where it was argued 
that the hopping parameter $\kappa_s = 0.13640$ used to generate the 
PACS-CS (2+1)-flavor gauge configurations generally leads to too large hadron masses, if they include a strange quark. Our findings confirm this picture. 

\begin{figure*}
\begin{center}
\centering
\includegraphics[width=7.8cm]{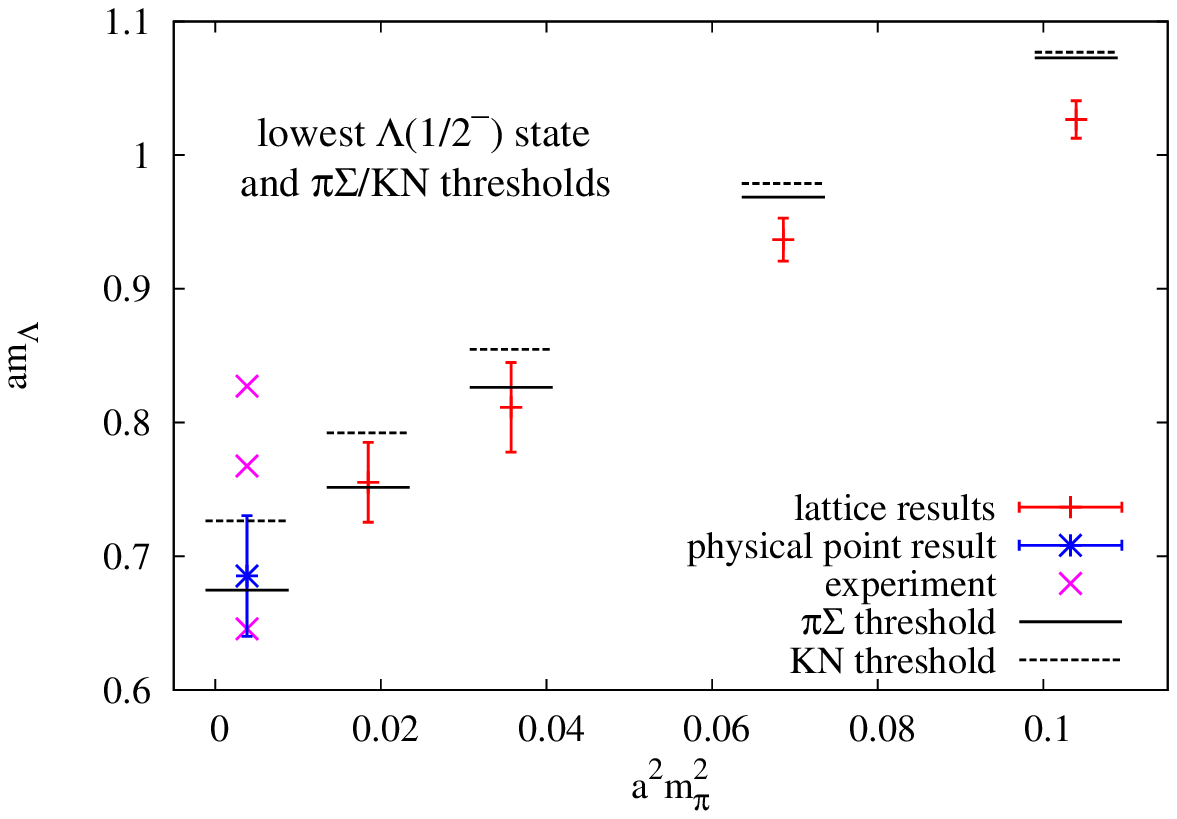}
\includegraphics[width=7.8cm]{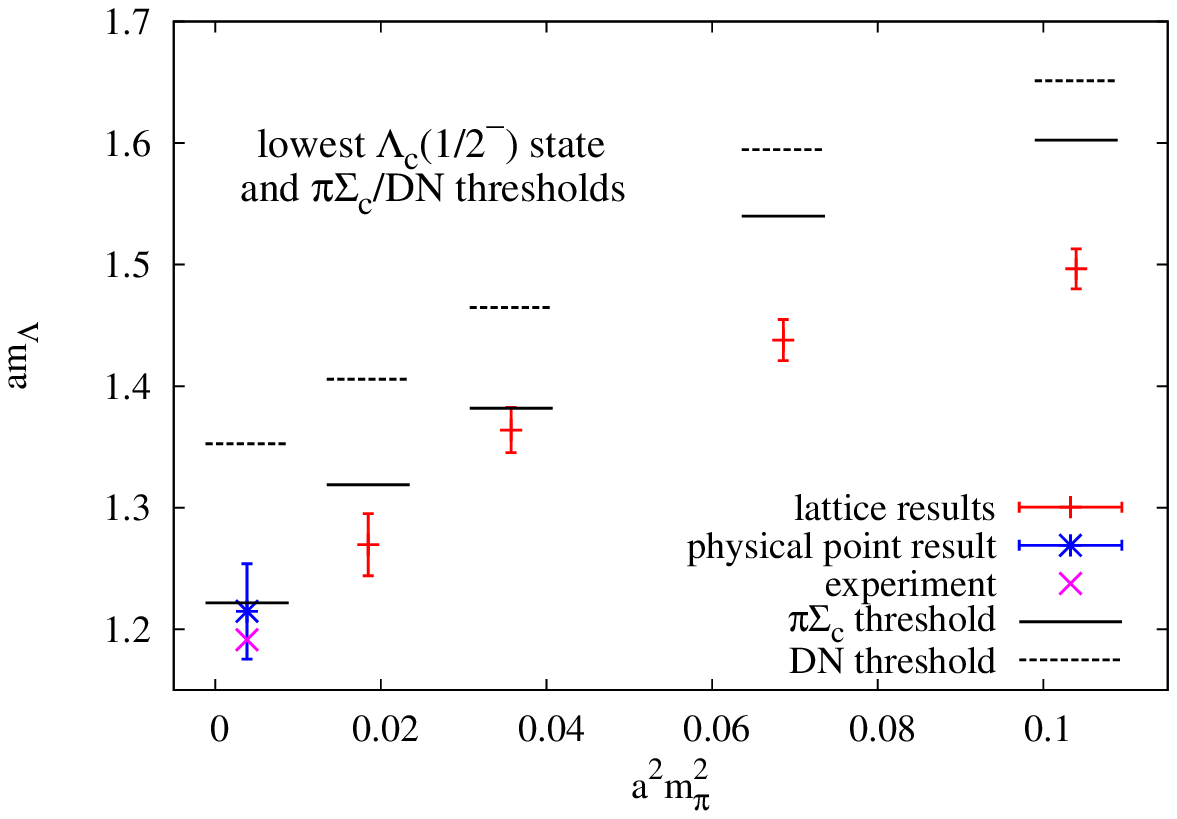}
\caption{Left plot: Mass of the lowest negative-parity $\Lambda$ baryon containing a strange quark, together with the $\pi \Sigma$ and $\overline{K}N$ 
thresholds, as a function of the squared pion mass. The threshold values are obtained from an independent analysis of single $\pi$, $\Sigma$, $K$ and $N$ 
masses. Their extrapolation to the physical point is performed by quadratically extrapolating these single masses individually and thereafter adding the extracted 
values. Right plot: Same as in left plot, but with $\Lambda_c$ instead of $\Lambda$ masses and $\pi \Sigma_c$ and $DN$ instead of $\pi \Sigma$ and $\overline{K}N$ 
thresholds. All results are shown in lattice units.}
\label{fig:thresholds}
\end{center}
\end{figure*}

Let us examine the lowest negative-parity state in some more detail, especially its relative position to  the $\pi \Sigma$ and $\overline{K}N$ 
thresholds, which is shown in the left plot of Fig.\,\ref{fig:thresholds}.  It is seen in this figure that for the heavier pion masses, the lowest $\Lambda(1/2^-)$ state lies 
below both thresholds 
and is therefore a bound state. As the pion mass is decreased to the lowest value studied in this work, however, it moves above the $\pi \Sigma$ threshold 
and hence turns into a resonance. Extrapolating both $\Lambda(1/2^-)$ mass and thresholds to the physical point, the order remains the same, 
with the $\Lambda(1/2^-)$ state lying between the $\pi \Sigma$ and $\overline{K}N$ thresholds, thus reproducing the level ordering observed in experiment. 
This is again consistent with results reported in earlier work \cite{Menadue:2011pd}. 
We note that we have found no evidence 
for any scattering state signal in a finite box, which 
should be the ``ground state" at the physical point.
As it was discussed recently in Ref.~\cite{Liu:2016wxq}, this absence of scattering states is likely 
due to the small overlap of our three-quark operators
with such scattering signals. 

Turning next to the excited states, it is seen that for positive parity, our lattice analysis is clearly not able to generate any states that could be related to the first- or second-excited 
state of the experimental spectrum. This feature has already been observed in an earlier study of two of the present authors \cite{Takahashi:2009bu}. 
On the other hand, for negative parity, our finding of two excited states lying close to the ground state qualitatively agrees with experiment 
(see the upper right plot in Fig. \ref{fig:extrapolation}). 

Next, we look at our results of the $\Lambda_c$ states. Here, much less in known from experiments, as for the relevant quantum numbers only the ground 
state has been found so far, while no established facts are available about possible excited states. From the lower two plots of Fig. \ref{fig:extrapolation}, we 
can, however, see that, for both positive and negative parity, these ground states are very well reproduced in our calculation. 
The extracted excited states are arranged like their strange counterparts: For positive parity the first excited state is found about 500 MeV above the ground state 
and for negative parity two excited states lie relatively close to the ground state. Especially for the negative-parity case, it is possible that such excited states will be 
found in future experimental searches, and it will be interesting to see whether our lattice QCD prediction can be verified in nature. 
In the right plot of Fig.\,\ref{fig:thresholds}, we furthermore show the position of the lowest $\Lambda_c(1/2^-)$ state in 
comparison with the $\pi \Sigma_c$ and $DN$ thresholds. We observe that the negative-parity $\Lambda_c$ baryon lies 
below the two thresholds for all pion masses but approaches the $\pi \Sigma_c$ threshold as the pion mass is 
decreased to the physical point. This is consistent with experiment, which finds the $\Lambda_c(1/2^-)$ mass right 
at the $\pi \Sigma_c$ threshold. 

Finally, here we briefly discuss effects of our employed finite lattice spacing and potential changes in our results 
in the continuum limit. We have performed our calculation with only a single lattice spacing, and it is therefore not 
possible to perform a reasonable extrapolation to the continuum limit. One can, however, try to roughly estimate this effect by 
consulting the available literature. 

We start first with the calculation dealing with the $\Lambda$ 
baryon containing only $u$, $d$ and $s$ quarks. Here, we can consult a similar work by two 
of the present authors \cite{Takahashi:2009bu}, in which the positive- and negative-parity $\Lambda$ baryon 
masses were studied for three different lattice spacings. As a result, it was found that the ground states for both 
parities do not strongly depend on the lattice spacing and therefore the effect of the 
continuum extrapolation can be expected to be small. Now, the lattice spacing used in the present 
work ($a = 0.0907$ fm) is even smaller than the ones used in Ref.\,\cite{Takahashi:2009bu}, which were 
all above 0.1 fm. Therefore, we do not expect the continuum limit to significantly alter our ground-state 
results. 
For the excited states, the work in Ref.\,\cite{Takahashi:2009bu}, however, obtained some rather large 
dependence on $a$, and we hence cannot exclude considerable systematic uncertainties due to the 
continuum extrapolation for these excited states. 

Next, we consider the potential continuum extrapolation effect for the $\Lambda_c$ states. In this case, 
one could expect to have a larger effect because of the large charm quark mass and the ensuing 
discretization errors of $\mathcal{O}(m_c a)$ of the clover action that we use. Here again, we can rely 
on a series of earlier work of two of the present authors \cite{Can:2012tx,Can:2013zpa,Can:2013tna}, 
in which various charmed hadrons have been studied with the clover action \cite{Can:2012tx,Can:2013zpa} 
and the Fermilab method \cite{Can:2013zpa,ElKhadra:1996mp} for which discretization errors are 
suppressed. In these works, $J/\psi$, $D$, $D^{\ast}$ and $\Xi_{cc}$ masses were computed with 
both actions, which allows us to get a rough estimate of the  $\mathcal{O}(m_c a)$ effects. Comparing 
the calculations, it is found that the results differ only by 2\,\%  or less, which gives an idea of the 
systematic discretization error effects caused by the charm quark. 

\subsection{Flavor decomposition}
In this section, we study the components of the eigenvectors obtained from our variational analysis of the correlation matrix. 
The interpolating fields are chosen such that they belong to either a singlet or an octet of the flavor $SU(3)$ group. 
From the couplings to the different interpolating fields, we can therefore make statements about the flavor structure of the extracted states. 

Let us first explain here our usage of the $SU(3)$ group terminology. When we discuss $\Lambda$ baryons with a strange quark, 
the flavor $SU(3)$ group has the conventional meaning, describing the symmetry of the three quark flavors $(u,d,s)$. 
When we switch to $\Lambda_c$ baryons, we make use of the same flavor $SU(3)$ group terminology, in which, however, the 
strange quark is now understood to be replaced by its charm counterpart: $(u,d,c)$. 
This allows us to study effects of the explicit flavor $SU(3)$ symmetry breaking as a function of the quark mass. 

Since in this work we are mainly interested in the decomposition of the states into flavor-singlet and flavor-octet components,  we 
combine the couplings to the three octet operators and compare their combined strength to the coupling of 
the singlet operator. For this purpose, we define
\begin{align}
g^{\bm{1}}_i &= \frac{|\Psi_{4i}|}{\sum_{I=1}^4 |\Psi_{Ii}|}, \label{eq:singletoctet1} \\
g^{\bm{8}}_i &= \frac{\sum_{J=1}^3 |\Psi_{Ji}|}{\sum_{I=1}^4 |\Psi_{Ii}|}, \label{eq:singletoctet2}
\end{align}
where $\Psi_{Ji}$ is given in Eq.\,(\ref{eq:overlap}). 
Note that $g^{\bm{1}}_i$ and $g^{\bm{8}}_i$ provide a quantitative estimate of the flavor-singlet and flavor-octet 
components of the state $i$. As in the present setting we are only able to investigate the relative coupling strengths, 
their sum is normalized to one. As in the effective mass plots in Fig. \ref{fig:eff.mass}, we compute the couplings at each time 
slice, define plateau regions, within which these couplings are approximately constant and determine our final numbers from a 
fit to the data points of the plateau. This procedure is repeated for all our hopping parameter combinations and finally the values 
are extrapolated to the physical $u$ and $d$ quark masses. The numerical results of the analysis are summarized in 
Tables \ref{tab:num.coupl.1} and \ref{tab:num.coupl.2} of Appendix A. 

\begin{figure*}
\begin{center}
\centering
\includegraphics[width=7.8cm]{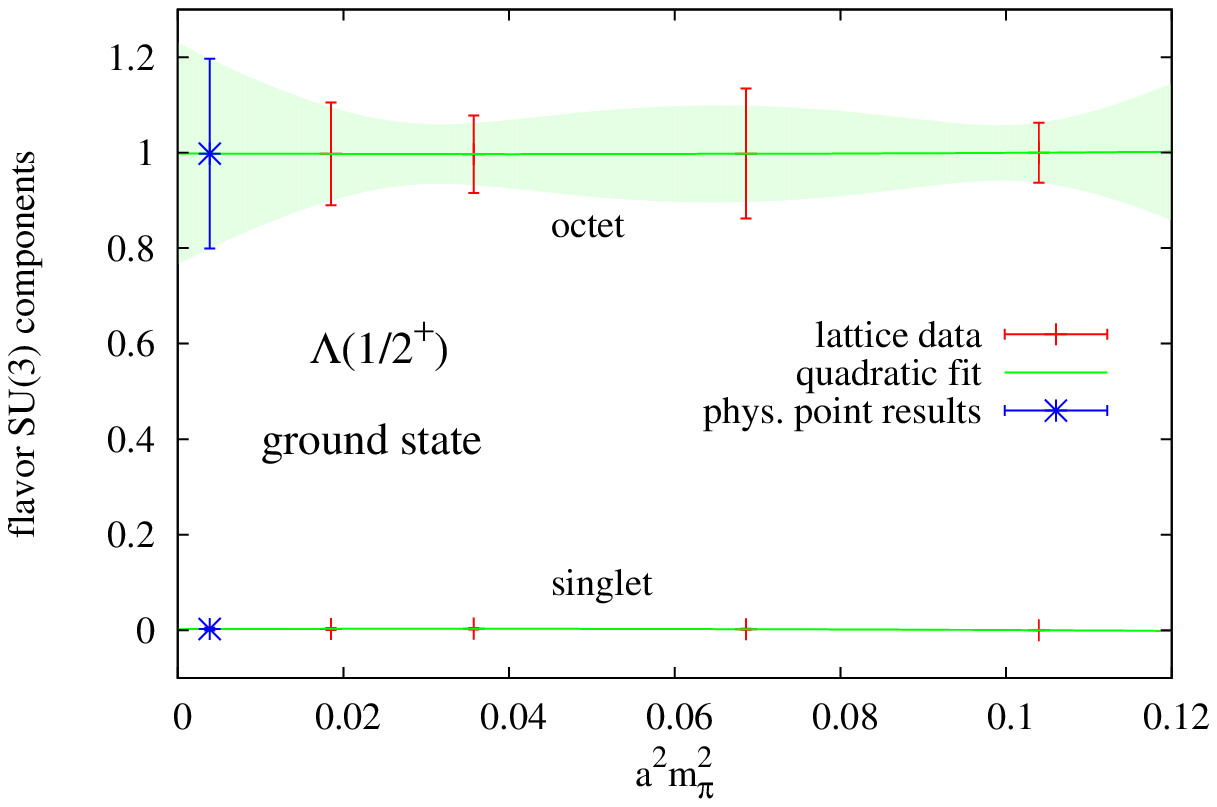}
\includegraphics[width=7.8cm]{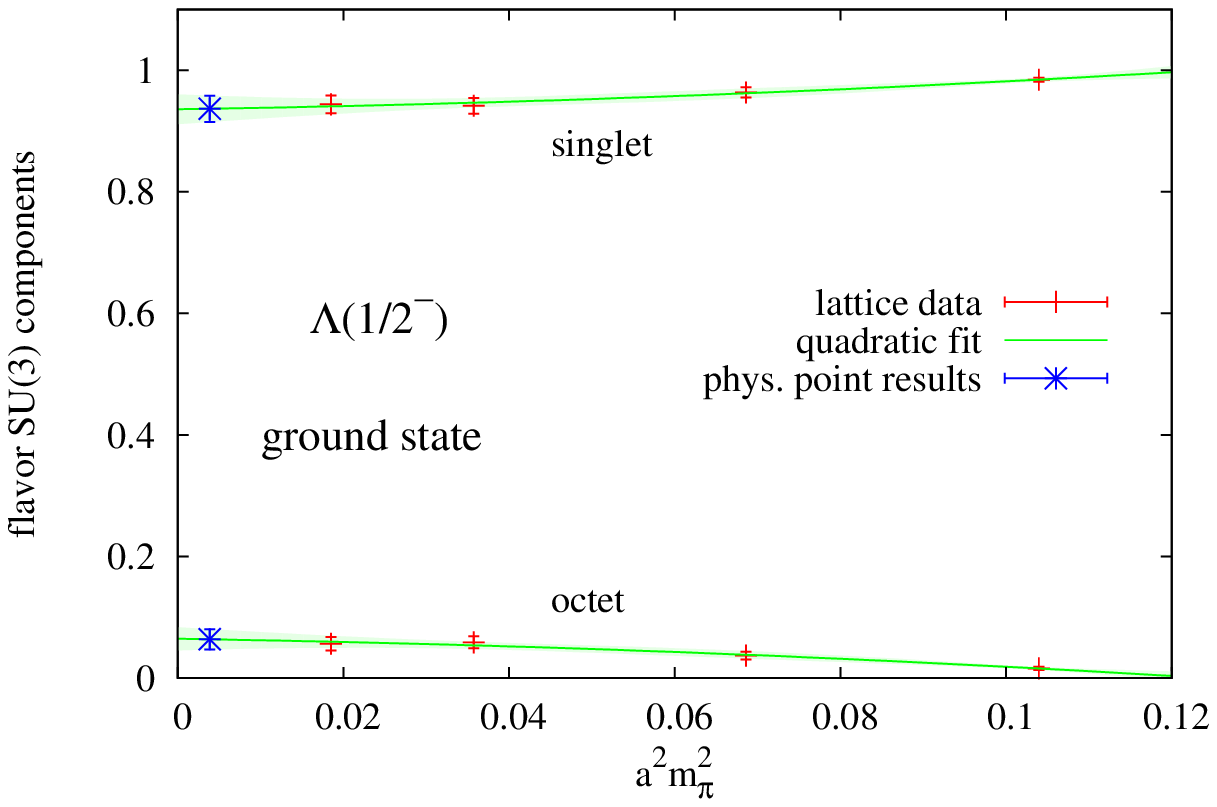}
\\
\includegraphics[width=7.8cm]{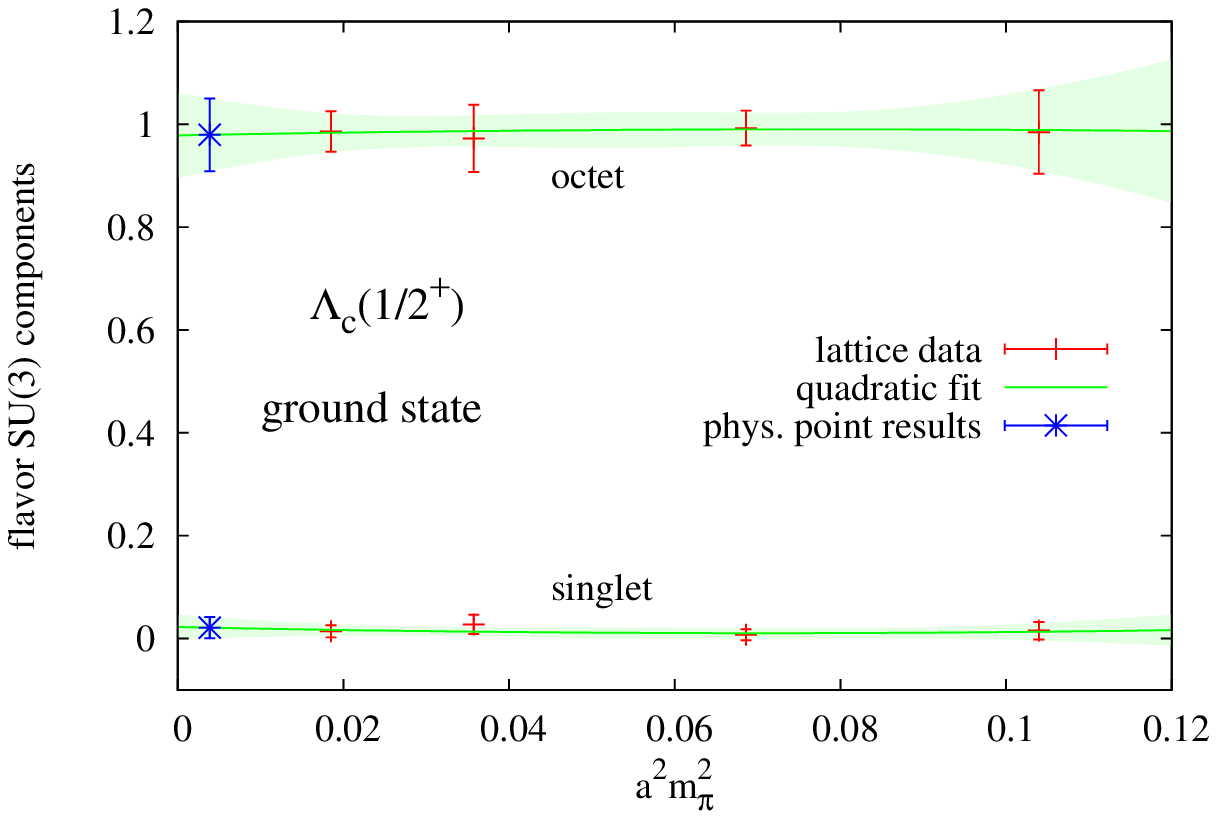}
\includegraphics[width=7.8cm]{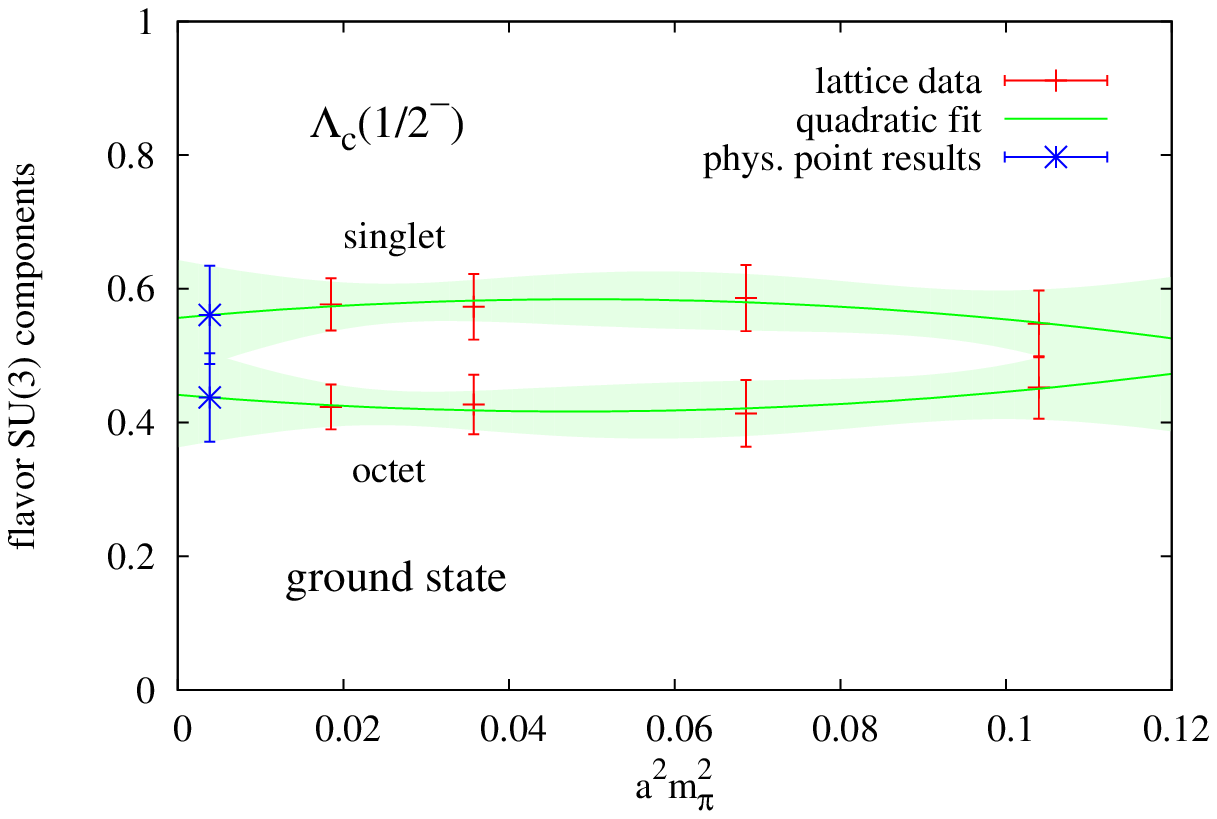}
\caption{
Normalized couplings of $\Lambda$ baryons of positive (left plots) and negative parity (right plots) containing a strange (upper plots) or charm quark (lower plots), 
shown as a function of the squared pion mass. The green curves are quadratic fits to the lattice points (shown in red). The blue points give the extrapolated physical point results. 
All plots are given in lattice units.}
\label{fig:coupl.extrap}
\end{center}
\end{figure*}

The behavior of the couplings as a function of $a^2 m_{\pi}^2$ is shown in Fig. \ref{fig:coupl.extrap} for the ground states of 
positive and negative parity and for the heavy quark hopping parameters $\kappa_s$ and $\kappa_c$ 
that correspond to the physical $s$ and $c$ quark masses. 
In these plots the quadratic extrapolations to the physical point are again indicated as green lines and shaded areas. 
Note that the flavor components do not strongly depend on $a^2 m_{\pi}^2$. Therefore, the chiral extrapolation 
to the physical point does not lead to a large systematic uncertainty, as can also be read off from Tables \ref{tab:num.coupl.1} and \ref{tab:num.coupl.1} 
of Appendix A, where both the results of linear and quadratic extrapolations are given, which all agree within their statistical errors. 
It is understood from these figures that the $\Lambda(1/2^+)$ ground state is clearly an octet-dominated state, with a singlet component 
too small to be visible in the plot. 
The situation is reversed for the $\Lambda(1/2^-)$ channel, whose ground state is dominantly flavor singlet, 
but has a somewhat larger octet component, which is increasing with a decreasing light quark mass. The growing subdominant component is a manifestation 
of the fact that, 
as we approach the physical $u$ and $d$ quark masses, the system is moved 
away from the flavor $SU(3)$ symmetric point, where the $u$, $d$ and $s$ quark masses are equal. 

For the flavor decomposition of the $\Lambda_c(1/2^+)$ ground state, the strong breaking of the  flavor $SU(3)$ symmetry due to the large 
$c$ quark mass has only a small effect, which means that 
this state remains clearly octet dominated. 
However, this result is different for $\Lambda_c(1/2^-)$, which has, in contrast to $\Lambda(1/2^-)$, equally strong components of both the singlet and 
octet. The flavor structure of the $\Lambda(1/2^-)$ ground state hence appears to be rather sensitive to the value of its heaviest valence quark. 
We consider potential implications of this finding for the structure of the physical $\Lambda(1405)$ and $\Lambda_c(2595)$ states in Sec. \ref{Discussion}. 

\begin{figure*}
\begin{center}
\centering
\includegraphics[width=7.8cm]{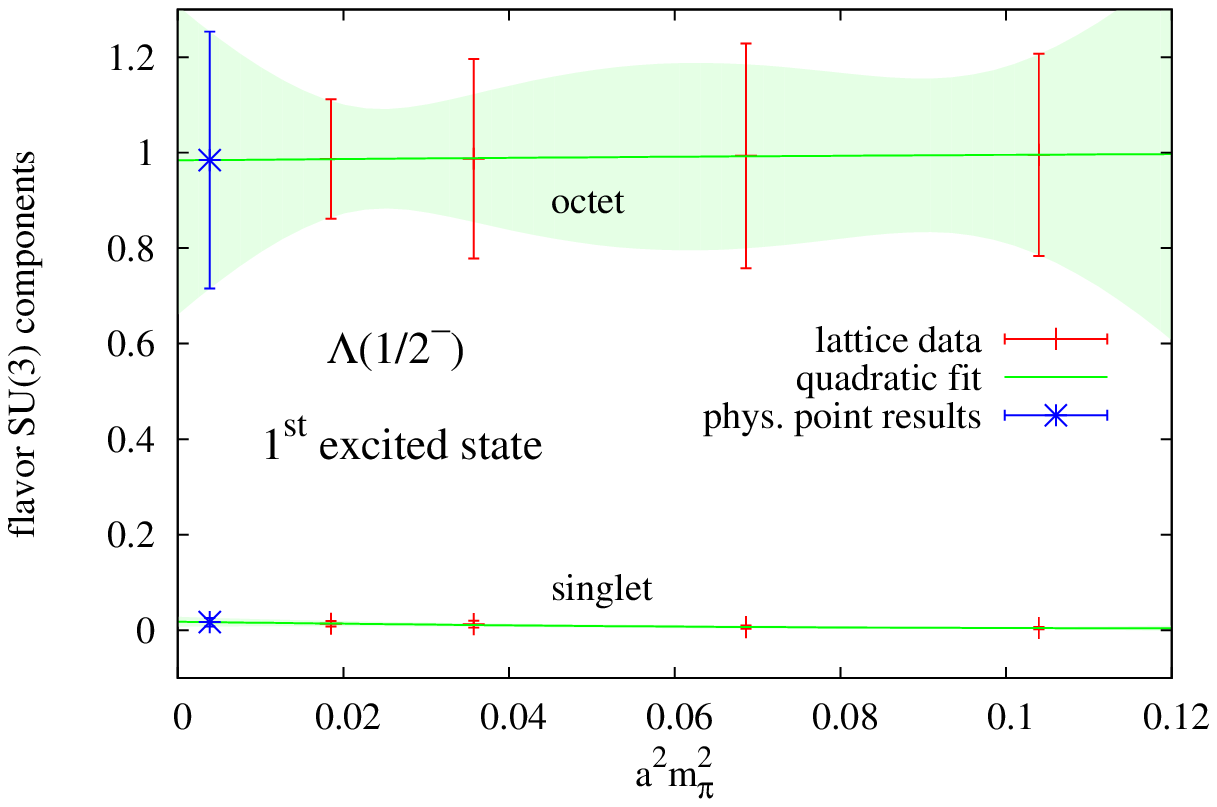}
\includegraphics[width=7.8cm]{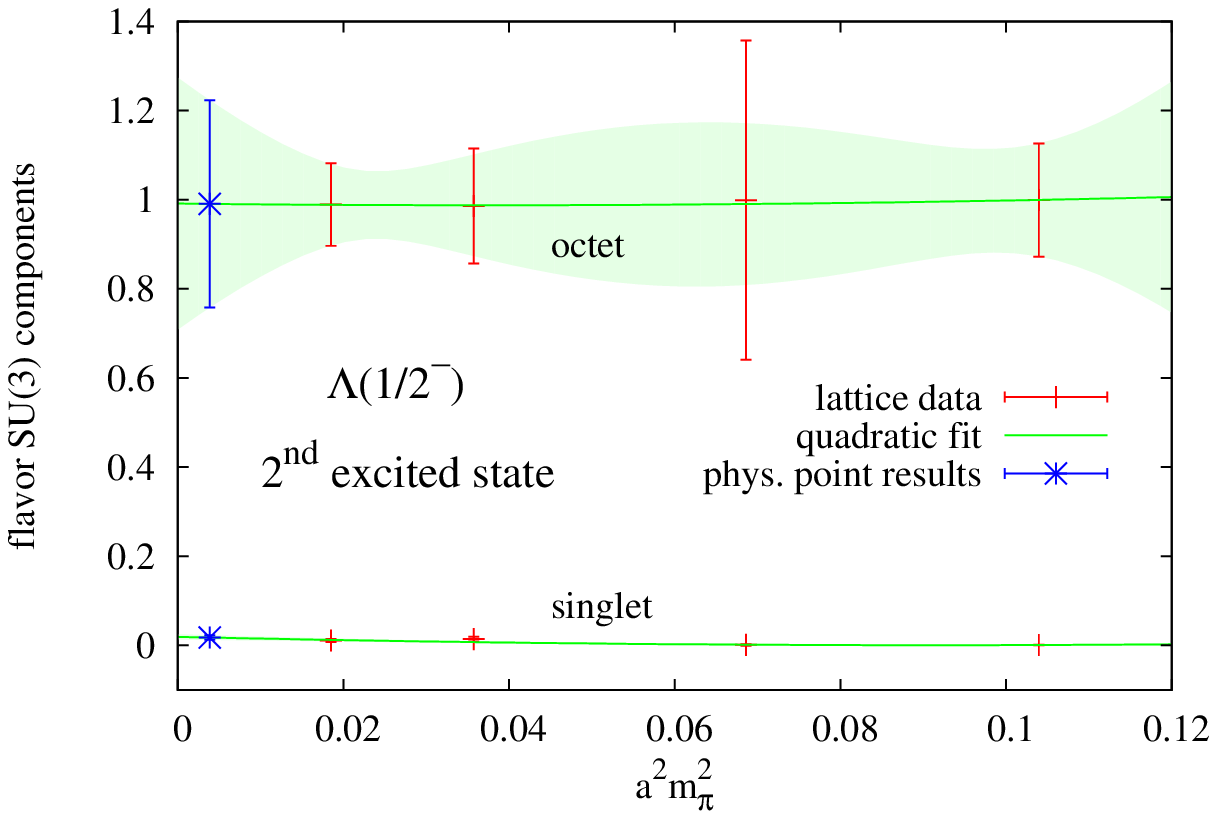}
\\
\includegraphics[width=7.8cm]{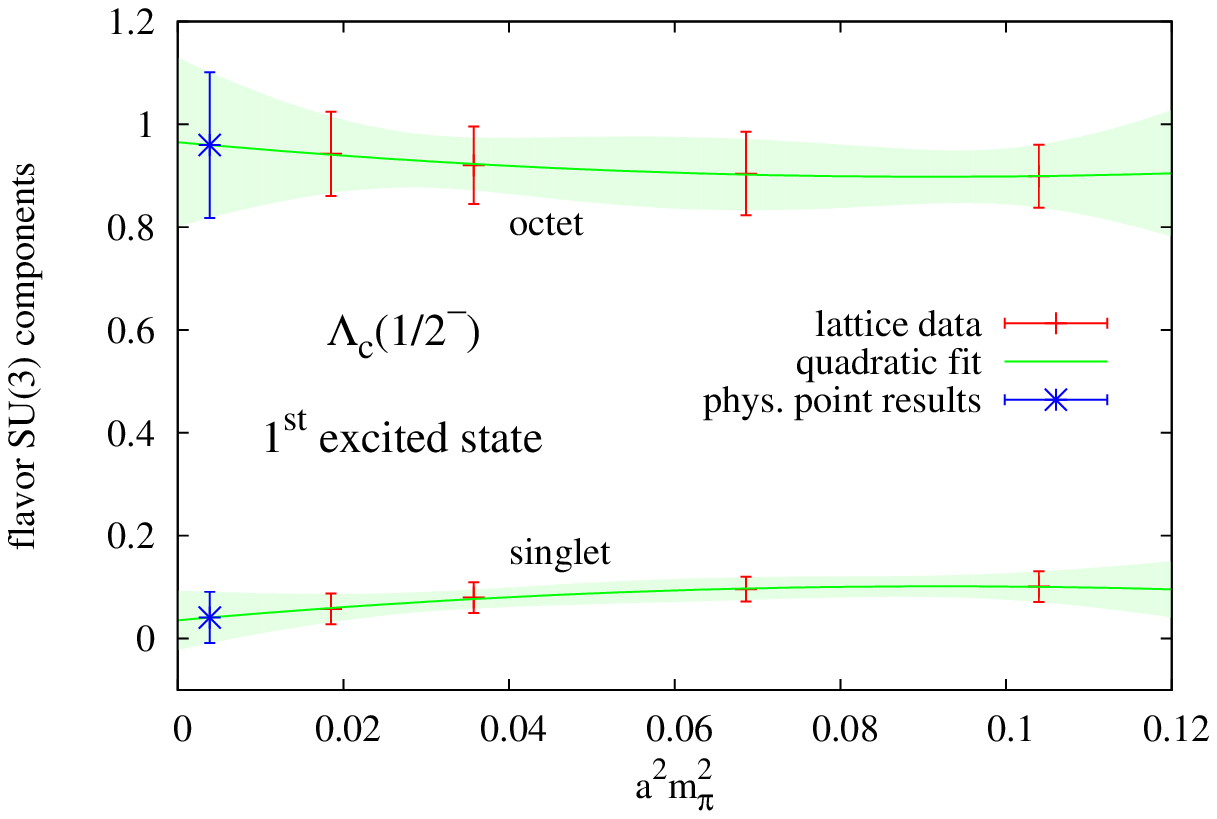}
\includegraphics[width=7.8cm]{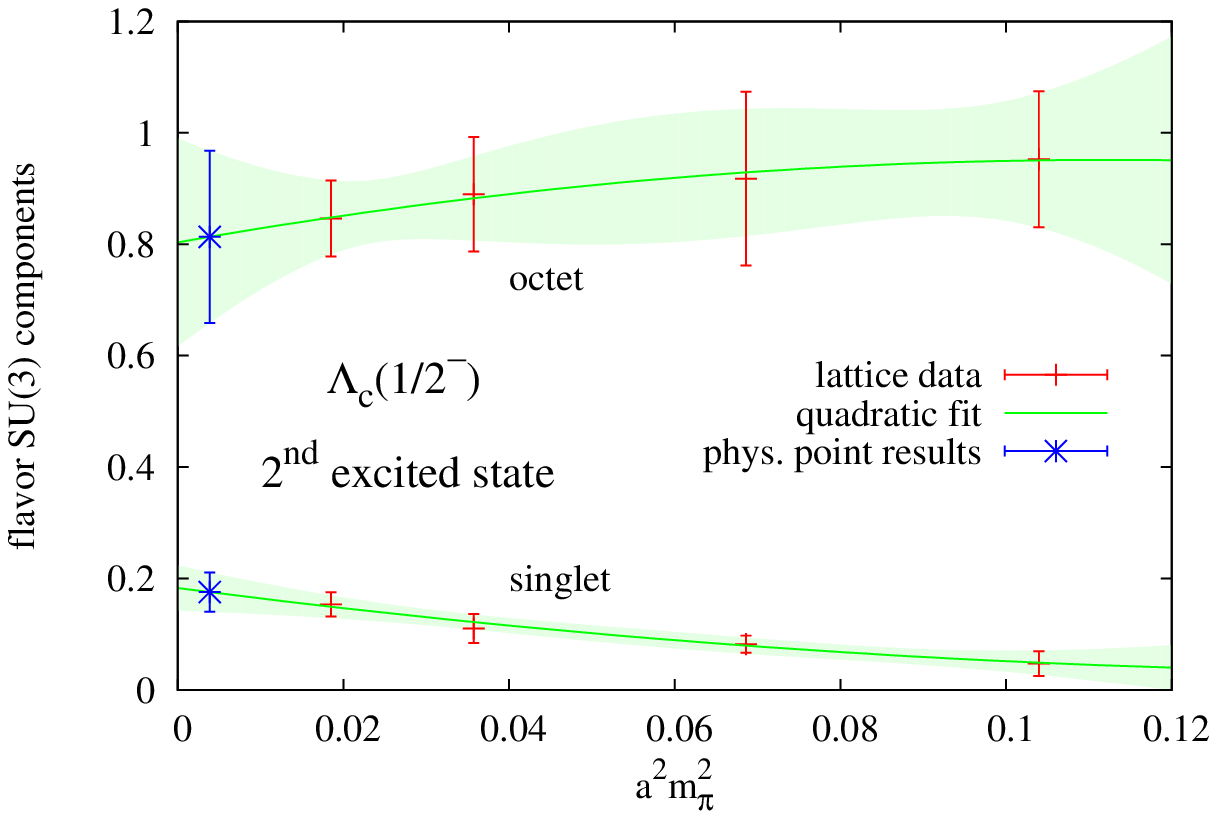}
\caption{
Normalized couplings of the first (left plots) and second (right plots) excited states of negative-parity $\Lambda$ baryons, containing 
a strange (upper plots) or charm quark (lower plots), shown as a function of the squared pion mass. The green curves are quadratic fits to the 
lattice points (shown in red). The blue points give the extrapolated physical point results. All plots are given in lattice units.}
\label{fig:coupl.extrap.exc}
\end{center}
\end{figure*}

As a last point, let us examine the flavor components of the negative-parity first and second excited states. Their 
extrapolations to physical point pion masses are shown in Fig.\,\ref{fig:coupl.extrap.exc}. 
As can be inferred from these plots, the first and second excited states for both $\Lambda(1/2^-)$ and $\Lambda_c(1/2^-)$ are predominantly flavor-octet states 
with small admixtures of singlet components. 
For $\Lambda$, both excitations are almost pure octet states, which agrees with the findings of \cite{Takahashi:2009bu,Menadue:2011pd,Engel}. 
The octet admixture is somewhat bigger for the $\Lambda_c$ state, especially for the second excited state, for which 
it reaches almost 20 \%. 

\subsection{Letting the $\Lambda$ evolve into $\Lambda_c$}
In the previous sections, we have concentrated our discussion on the physical states corresponding to the hopping parameters $\kappa_s$ 
and $\kappa_c$ for the heaviest valence quark. 
Here, we study how the two states evolve into one another as the hopping parameter is varied from $\kappa_s$ to $\kappa_c$. 
For this purpose, we have calculated the masses and couplings as shown in the two previous sections for three more hopping parameters 
that lie between $\kappa_s$ and $\kappa_c$. The numerical results are given in Tables 
\ref{tab:num.eff.masses.2}, \ref{tab:num.eff.masses.3}, 
\ref{tab:num.coupl.1} and \ref{tab:num.coupl.2} 
of Appendix A. 

\begin{figure*}
\begin{center}
\centering
\includegraphics[width=12.0cm]{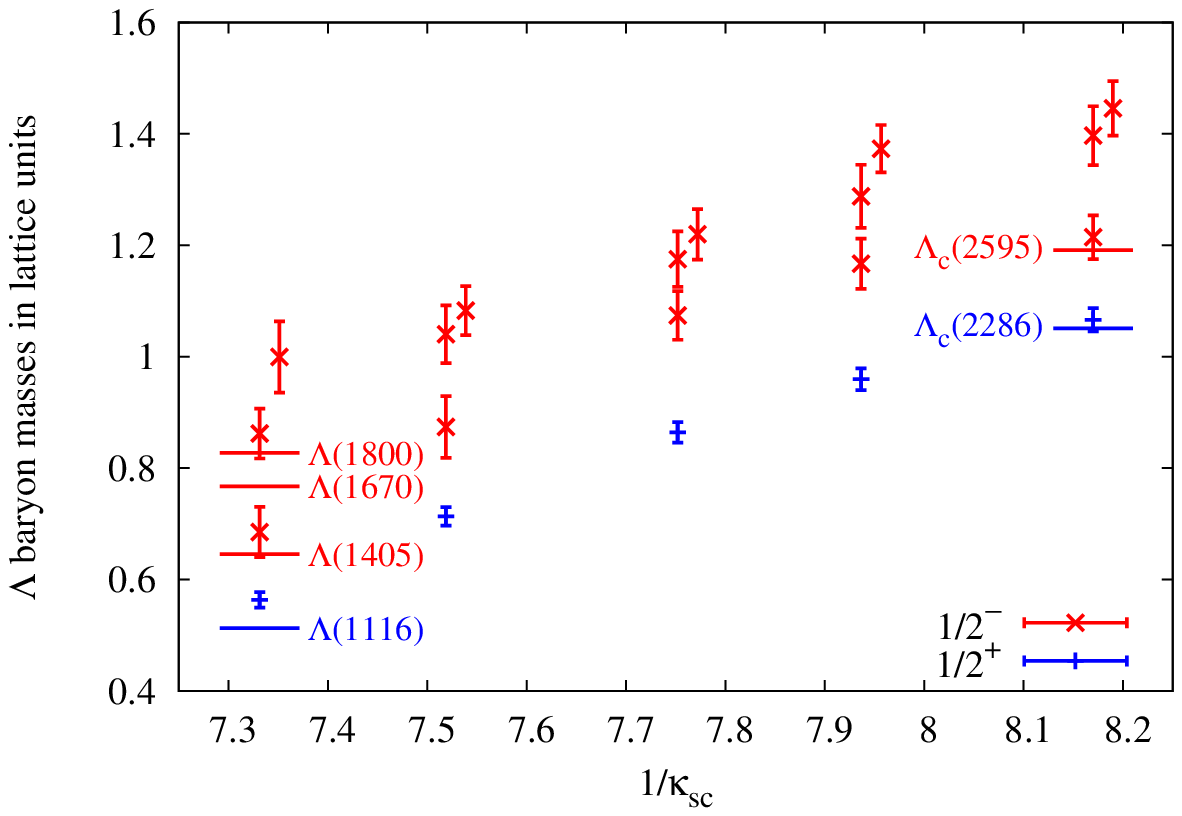}
\caption{
Masses of $\Lambda$ baryon ground states of positive and negative parity, 
shown as a function of $1/\kappa_{sc}$, $\kappa_{sc}$ being the hopping parameter of the heaviest 
valence quark. 
For $\Lambda(1/2^-)$ states, the first and second excited states are also shown. 
For better visibility, the position of the data points of the second excited state is shifted slightly to 
the right. 
The plot is given in lattice units.
}
\label{fig:mass.evolve}
\end{center}
\end{figure*}

Let us first study how the hadron masses behave as a function of $1/\kappa_{sc}$. In Fig.~\ref{fig:mass.evolve}, 
the positive-parity ground state and the lowest three states of negative parity are shown, which have been extrapolated to 
physical $u$ and $d$ quark masses. 
\begin{figure*}
\begin{center}
\centering
\includegraphics[width=7.8cm]{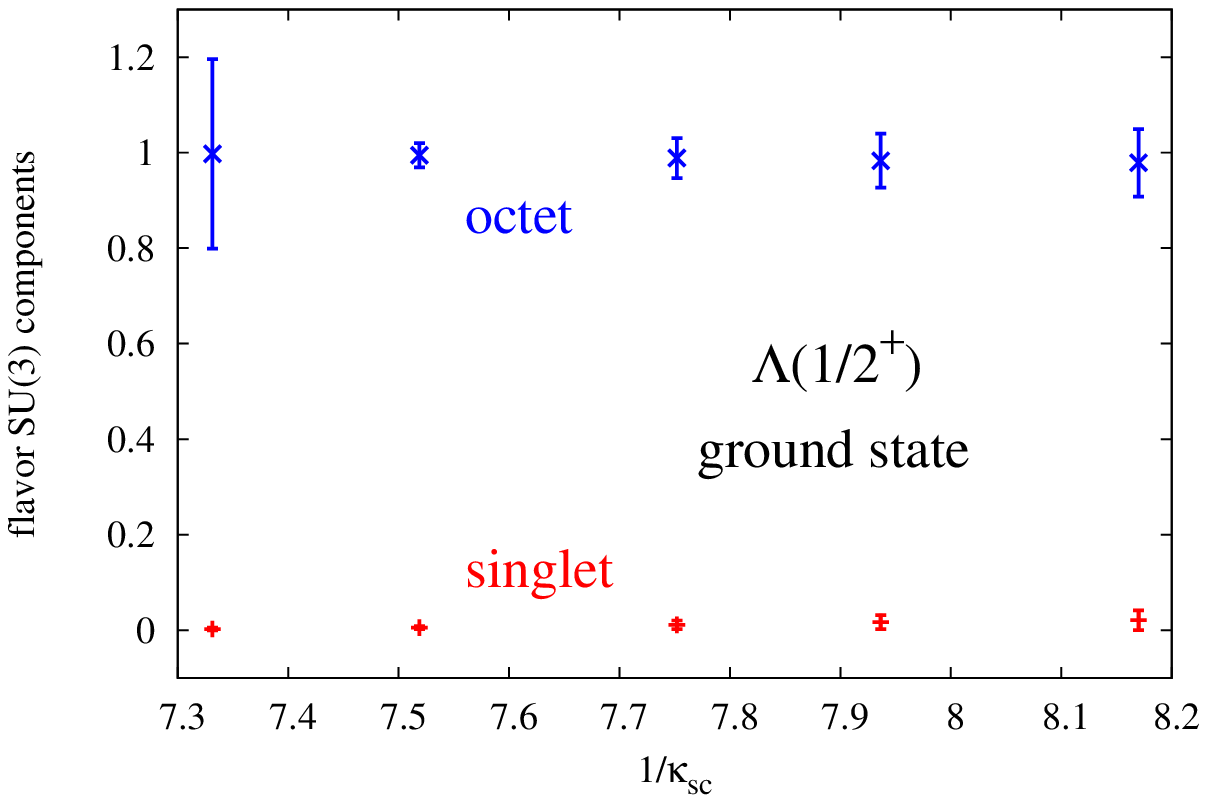}
\includegraphics[width=7.8cm]{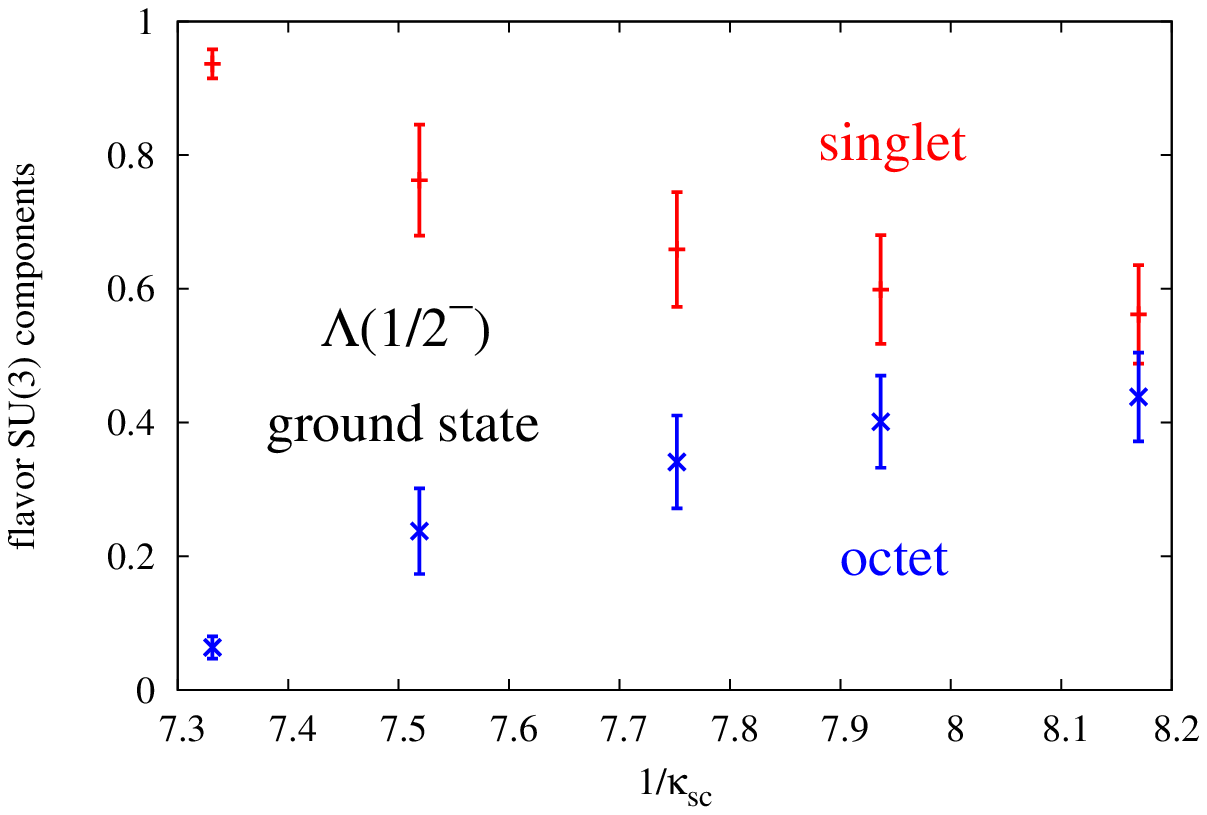}
\caption{
Coupling components of singlet and octet operators 
for positive-parity (left plot) and negative-parity (right plot) $\Lambda$ baryons 
shown as a function of $1/\kappa_{sc}$, with $\kappa_{sc}$ being the hopping parameter of the heaviest valence quark. 
Both plots are given in lattice units.
}
\label{fig:coupl.evolve}
\end{center}
\end{figure*} 
\begin{figure*}
\begin{center}
\centering
\includegraphics[width=7.8cm]{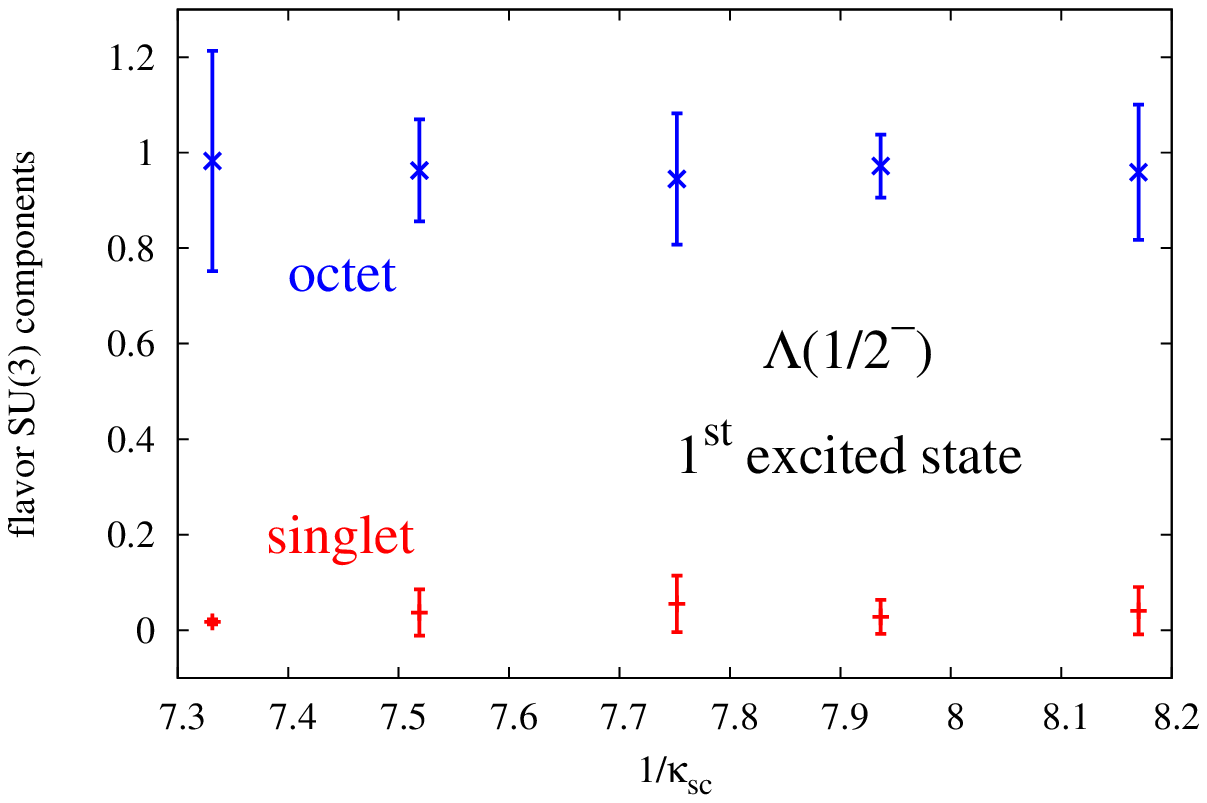}
\includegraphics[width=7.8cm]{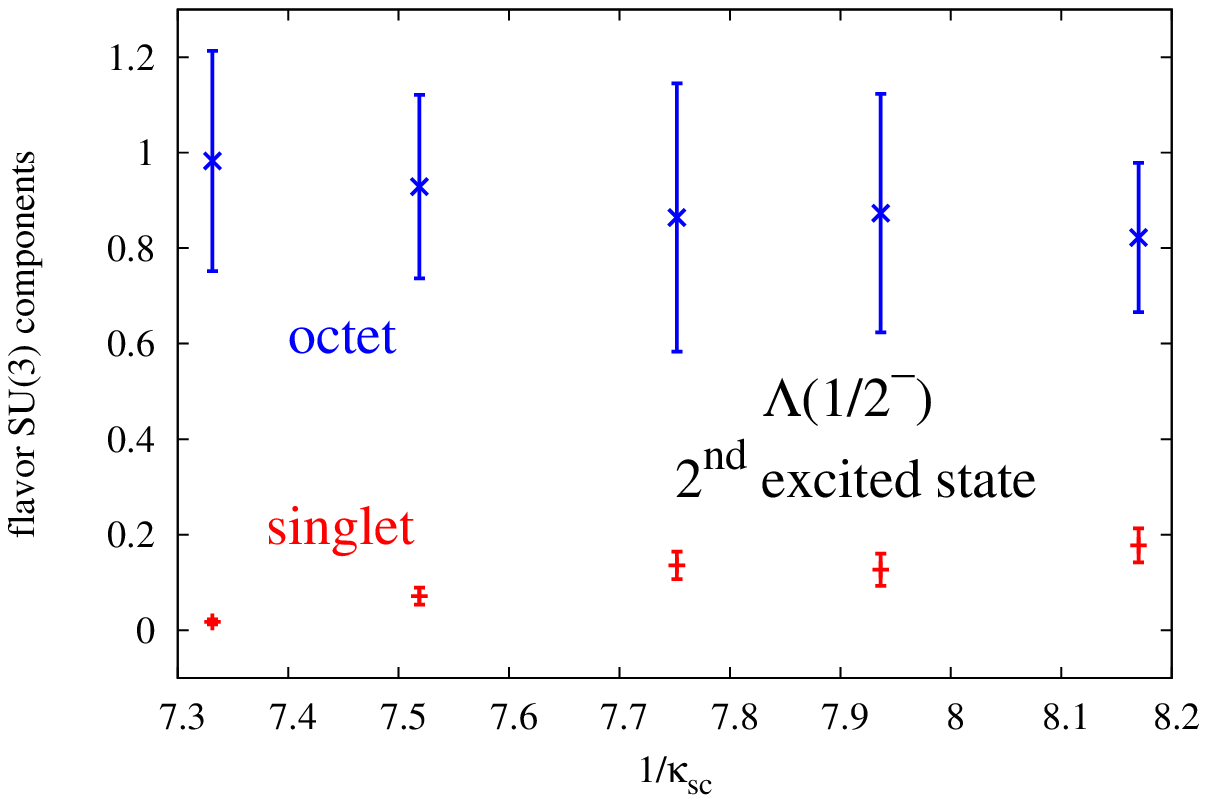}
\caption{
Coupling components of singlet and octet operators of negative-parity $\Lambda$ baryon first (left plot) and second (right plot) excited states,   
shown as a function of $1/\kappa_{sc}$, $\kappa_{sc}$ being the hopping parameter of the heaviest valence quark. 
Both plots are given in lattice units.
}
\label{fig:coupl.evolve.exc}
\end{center}
\end{figure*} 
It is seen in this figure that the masses grow smoothly (and almost linearly) with increasing $1/\kappa_{sc}$. 
The excited states tend to have larger errors but otherwise show essentially the same 
monotonously increasing behavior. 

Next, we investigate how the flavor structure of the $\Lambda$ states evolve as a function of $1/\kappa_{sc}$, 
focusing first on the ground states. Their couplings to singlet and 
octet operators are shown in Fig.\,\ref{fig:coupl.evolve}, where, as above, the extrapolated physical point values were used. 
As one could anticipate already from the left figures of Fig.\,\ref{fig:coupl.extrap}, the singlet and octet components of the positive-parity 
ground state depend only weakly on $1/\kappa_{sc}$, which is seen in the left plot of Fig.\,\ref{fig:coupl.evolve}. 
The situation is quite different for negative parity, for which both ground-state components exhibit a strong dependence on the 
value of the hopping parameter. As can be inferred from the right plot of Fig.\,\ref{fig:coupl.evolve}, the initially singlet-dominated 
state evolves with increasing quark mass into a state with approximately equal strength of singlet and octet components. This 
observation indicates that the physical states $\Lambda(1405)$ and $\Lambda_c(2595)$ have a different 
internal structure and that, in particular, the properties of the $\Lambda(1405)$ are closely related to the specific value of 
the physical strange quark mass. 

For the negative-parity excited states, we observe a behavior that is different from the ground state. 
As shown in Fig.\,\ref{fig:coupl.evolve.exc}, 
the first excited state remains an almost pure octet for all hopping parameters that we have studied in this work. 
For the second excited state, the singlet component exhibits a small enhancement as the quark mass is 
increased from $m_s$ to $m_c$, remaining, however, below 20\,\%.  

It is interesting to see that these excited states do not 
change their flavor structure much with increasing quark mass, which 
further accentuates the observation that the $\Lambda(1/2^-)$ ground state indeed appears to be quite peculiar with 
regard to its flavor decomposition. 

In relation to the contents of this section, a short comment about the partial quenching effect is in order. 
Our calculations  are indeed partially quenched because the sea quarks always remain $u$, $d$ and $s$, while the 
strange valence quark is gradually shifted to charm. 
Let us try to give a plausible assessment of these effects by first focusing 
on the two ``physical" points, $\Lambda$ and $\Lambda_c$. 
For $\Lambda$, where the heavy quark is the strange quark, the valence quark hopping parameters 
agree with those of the gauge configurations for all $u$, $d$ and $s$, and there is thus no issue with 
partial quenching. 
For the $\Lambda_c$ case, we have $u$, $d$ and $s$ sea quarks, while the valence quarks consist of 
$u$, $d$ and $c$. Therefore, here we are simply neglecting dynamical charm quarks, which does not seem to 
be a very problematic approximation as charm quarks are quite a bit heavier than the typical QCD scales. 
Thus, in the $\Lambda_c$ limit, partial quenching should not cause any effects that are too strong either. 
Now, between these two limits, partial quenching could have some notable effect, and the middle three 
data points in Figs.\,\ref{fig:mass.evolve}-\ref{fig:coupl.evolve.exc} could indeed be modified once partial quenching is removed. 
It is, however, very unlikely that the qualitative behavior displayed in these figures is strongly modified in any way 
as the two limiting points are practically fixed. 
Our conclusions, especially about the flavor structure of the $\Lambda$ baryons and their modification 
as the strange quark is changed to charm, are therefore not expected to be affected by partial quenching. 

\section{\label{Discussion} Discussion}
The flavor structure of the $\Lambda$ baryons, clarified
by changing the heavy quark mass from strange to charm, 
shows that
the $SU(3)$ classification works well for $uds$-$\Lambda$ baryons
but not for $udc$-$\Lambda$:
When the heavy quark's mass is as light as the strange, all the states are classified into either pure singlet or pure octet states 
with little contamination by other representations.
On the other hand, when the heavy quark mass is gradually raised,
the $SU(3)$ classification breaks and states are described
by the admixture of singlet and octet components.
In the heavy quark limit, $1/\kappa_{sc}\rightarrow\infty$,
the spin of light and heavy quarks decouples 
and the heavy quark symmetry will become an exact symmetry 
of the system. 
In order to get a deeper insight on the $\Lambda$'s structure, 
we discuss how our results can be interpreted in terms of the internal structure of the $\Lambda$ states. 
For this purpose, it is useful to briefly remind the reader of the concept of diquarks and their most relevant excitation 
modes, which will become crucial especially for discussing the $\Lambda_c$ states, for which the heavy charm quark is 
expected to play the role of a static color source; hence, the lowest few excitations should be dominated by 
the dynamics of the remaining light diquark system. 
In this section we focus on the negative-parity states with total spin $1/2$. 

Let us therefore, for a moment, discuss a simple nonrelativistic three-quark model, which will help us to 
understand the basic properties of the diquark excitations. 
Here, we should emphasize that it is not our purpose to discuss the quark model on the same level as our obtained lattice QCD results. 
The quark model merely serves as a guideline for potentially interpreting the lattice findings in terms of constituent quark degrees of 
freedom. 
We assume the masses of two quarks to be equal and light ($m_q$) and one to be heavy ($m_Q$). 
Using a confining harmonic oscillator potential and two internal coordinates 
$\bm{\rho} = \bm{r}_{q_2} - \bm{r}_{q_1}$, $\bm{\lambda} = \bm{r}_Q - \frac{1}{2}(\bm{r}_{q_2} + \bm{r}_{q_1})$ 
with their respective conjugate momenta $\bm{p}_{\rho}$, $\bm{p}_{\lambda}$, the Hamiltonian of this system can be 
straightforwardly written down as 
\begin{equation}
\begin{split}
H =&~ \sum_{i} \frac{\bm{p}_i^2}{2 m_i} + \sum_{i<j} \frac{3k}{2} (\bm{r}_i - \bm{r}_j)^2 \\
=&~~ \frac{\bm{p}_{\rho}^2}{2m_{\rho}} + \frac{\bm{p}_{\lambda}^2}{2m_{\lambda}} 
+ \frac{1}{2} m_{\rho} \omega^2_{\rho} \bm{\rho}^2 + \frac{1}{2} m_{\lambda} \omega^2_{\lambda} \bm{\lambda}^2, 
\end{split}
\end{equation}
with the reduced masses 
\begin{equation}
m_{\rho} = \frac{1}{2} m_q, \hspace{0.5cm} m_{\lambda} = \frac{2 m_q m_Q}{2 m_q + m_Q}.  
\end{equation}
Most importantly, the ratio of the excitation energies of the two modes appearing in this model, $\omega_{\rho}$ and  
$\omega_{\lambda}$, can be given as 
\begin{equation}
\frac{\omega_{\lambda}}{\omega_{\rho}} = \sqrt{\frac{1}{3} \big(1 + 2m_q/m_Q \big)} < 1, \hspace{0.5cm}(m_Q > m_q), 
\label{eq:ex.energy}
\end{equation}
which shows that as long as $m_Q$ is larger than $m_q$, the lowest excited state will be a $\lambda$ mode, that is, 
an excitation of the center-of-mass motion of the two light quarks with respect to the heavy quark. The next 
energy level should then be a $\rho$ mode, which is an excitation of the relative motion of the two light quarks.  

It is instructive to study the wave functions of the $\lambda$ and $\rho$ modes with
respect to their $SU(3)$ flavor structure. 
Here, we only mention the decomposition of the wave functions in terms of their flavor-singlet and 
flavor-octet components and refer the interested reader to Ref.~\cite{Yoshida} for more details. 
Taking into account the spin degrees of freedom, there are two possible combinations for the $\rho$ mode 
and one for the $\lambda$ mode: 
\begin{align}
|\Lambda; \rho(1/2) \rangle &= \frac{1}{\sqrt{2}} \Big(|\Lambda; \bm{8}(1/2) \rangle - |\Lambda; \bm{1}(1/2) \rangle \Big), \label{eq:wave.func1} \\
|\Lambda; \lambda(1/2) \rangle &= \frac{1}{\sqrt{2}} \Big(|\Lambda; \bm{8}(1/2) \rangle + |\Lambda; \bm{1}(1/2) \rangle \Big), \label{eq:wave.func2} \\
|\Lambda; \rho(3/2) \rangle &= |\Lambda; \bm{8}(3/2) \rangle. \label{eq:wave.func3}
\end{align} 
Here, the numbers in brackets stand for the total spin of the three quarks, 
which can be $1/2$ or $3/2$ before it is combined with the orbital angular momentum of spin 1. For both $\rho$-mode 
combinations of Eqs.\,(\ref{eq:wave.func1}) and (\ref{eq:wave.func3}), the two light quarks are in a spin-1 state, while 
for the $\lambda$ mode of Eq.\,(\ref{eq:wave.func2}), it is in a spin-0 state. From this decomposition it can be seen that 
the $\lambda$ mode \textit{must} have flavor-singlet and flavor-octet components of the same size, while the $\rho$ mode can be 
either an equally mixed singlet and octet state of Eq.\,(\ref{eq:wave.func1}) or a pure octet state of Eq.\,(\ref{eq:wave.func3}) 
(or a mixture of the two). 

Let us check if and how our lattice QCD results can be understood and interpreted with the help of the above simple quark-model 
considerations. Looking first at the lowest $\Lambda_c(1/2^-)$ state, one notes that for this state the lattice findings 
almost perfectly match with the quark-model predictions. According to Eq.(\ref{eq:ex.energy}), the lowest excitation should 
be a $\lambda$ mode, which from Eq.(\ref{eq:wave.func2}) must have equal magnitudes of singlet and octet components. 
The results shown in the bottom right plot of Fig.\,\ref{fig:coupl.extrap} agree with this picture, which is a strong 
indication that this state indeed represents a $\lambda$ mode. 
To further confirm this finding, we have studied the relative sign of the individual couplings to the singlet and 
octet operators and have found that it agrees with that of the $\lambda$-mode state of Eq.\,(\ref{eq:wave.func2}). 

Remembering the right plot of Fig.\,\ref{fig:coupl.evolve}, we observe that such a quark-model-type 
interpretation only holds for a sufficiently heavy quark mass $m_c$, as the lowest $\Lambda(1/2^-)$ is 
rather a singlet-dominated state, which 
is a consequence of the still unbroken flavor symmetry and 
cannot be easily explained in a simple three-quark model. 
This is in agreement with the recent lattice QCD study of Hall \textit{et al.} \cite{Hall}, which found evidence that this 
state is dominantly a $\overline{K}N$ molecule. 
In this sense, Fig.\,\ref{fig:coupl.evolve} demonstrates how the diquark degrees of freedom gradually emerge as the 
heavy quark mass in the $\Lambda$ system is shifted from $m_s$ to $m_c$. 

Next, we examine the second and third $\Lambda_c(1/2^-)$ states, 
which are considered to be $\rho$ modes. 
As can be seen in Fig.\,\ref{fig:coupl.extrap.exc}, the first (second) excited state is octet dominated with a singlet admixture 
of about 10\,\% (20\,\%). 
One may naively think that the SU(3) symmetry appears to hold, and a simple quark-model interpretation is not suitable for these states,
since according to Eqs.~(\ref{eq:wave.func1}) and (\ref{eq:wave.func3}), 
one $\rho$ mode should be octet dominant and the other should be
an equal admixture of octet and singlet components.
In reality, however, these two $\rho$ modes mix with each other,
as their quantum numbers are the same. 
Qualitatively, 
these states can be understood by assuming that the two states are pure 
eigenstates of the total spin of the two light quarks (called $j$). In the heavy quark limit, the heavy quark spin decouples, and hence $j$ becomes 
a good quantum number. The two $\rho$-mode states of Eqs.\,(\ref{eq:wave.func1}) and (\ref{eq:wave.func3}) can be 
decomposed into states of fixed $j$ as given below \cite{Yoshida}: 
\begin{align}
|\Lambda; \rho(1/2) \rangle &= \sqrt{\frac{1}{3}}|\Lambda; j=0 \rangle - \sqrt{\frac{2}{3}}|\Lambda; j=1 \rangle, \label{eq:wave.func4} \\
|\Lambda; \rho(3/2) \rangle &= \sqrt{\frac{2}{3}}|\Lambda; j=0 \rangle + \sqrt{\frac{1}{3}}|\Lambda; j=1 \rangle. \label{eq:wave.func5}
\end{align} 
Using the above two equations in combination with Eqs.\,(\ref{eq:wave.func1}) and (\ref{eq:wave.func3}), we get 
\begin{align}
|\Lambda; j=0 \rangle =&~ -\sqrt{\frac{1}{6}}|\Lambda; \bm{8}(1/2) \rangle + \sqrt{\frac{2}{3}}|\Lambda; \bm{8}(3/2) \rangle \nonumber \\  
&~+ \sqrt{\frac{1}{6}}|\Lambda; \bm{1}(1/2) \rangle, \label{eq:wave.func6} \\
|\Lambda; j=1 \rangle =&~ \sqrt{\frac{1}{3}}|\Lambda; \bm{8}(1/2) \rangle + \sqrt{\frac{1}{3}}|\Lambda; \bm{8}(3/2) \rangle \nonumber \\
&~- \sqrt{\frac{1}{3}}|\Lambda; \bm{1}(1/2) \rangle. \label{eq:wave.func7}
\end{align}
If we now examine the flavor components of these states, we see that both of them are octet dominated, 
which qualitatively agrees with our lattice results. 

Naturally, the agreement is not perfect, for which there can be multiple causes. For example, for physical charm quark 
masses, $j$ is not a good quantum number, and the energy eigenstates are hence mixed, in reality. Quark-model calculations show that the first (second) 
excited $\Lambda_c(1/2^-)$ state is indeed a $j=0$ ($j=1$) dominated state, with a respective minor spin component of about 20\,\% (for both 
$j=0$ and $j=1$) \cite{Yoshida}. 

\section{\label{SummaryConclusion} Summary and Conclusion}
In this work we have studied $\Lambda$ baryons containing either an $s$ or a $c$ quark and have examined how their masses and 
flavor structures change as the mass of the heaviest valence quark is gradually increased from $s$ to $c$. 
We have investigated states of both positive and negative parity and spin 1/2. For these states, we have not only 
studied the ground state but also the first few excited states. 
The behavior of the $\Lambda$ baryon masses as a function of the heavy quark mass is shown in Fig.\,\ref{fig:mass.evolve}, 
where one observes a smooth and almost linear behavior of the energy levels, while their relative energy differences and level orderings 
remain effectively constant. One also sees that while for the $\Lambda$ states with an $s$ quark, our extracted masses lie consistently 
above the experimental values, the agreement between our calculation 
and experiment is excellent for all known $\Lambda_c$ states. 

The chirally extrapolated $SU(3)$ flavor components of the positive-parity ground state and the lowest three negative-parity 
states are shown in Figs.\,\ref{fig:coupl.evolve} and \ref{fig:coupl.evolve.exc}. Somewhat surprisingly, we find that for almost 
all states, the flavor decomposition remains approximately constant as $s$ is changed to $c$. The notable exception is the lowest 
$\Lambda(1/2^-)$ state, which changes from singlet dominated to an equal mixture of singlet and octet components. 

Finally, in an attempt to provide an intuitive physical picture for the above findings, we have discussed a simple quark model with three 
basic valence quarks and have examined whether it can explain the features of the extracted spectrum. We have especially focused on 
the possible interpretation of the negative-parity states as $\lambda$ modes or $\rho$ modes, which are diquark excitations of the 
$u$ and $d$ quarks. As a result, we found that for the negative-parity $\Lambda$ states the quark-model description does not appear to 
be appropriate and thus should be interpreted 
by means of other degrees of freedom (such as mesons and baryons). 
On the other hand, the quark model is fairly successful for the negative-parity $\Lambda_c$ states. Namely, the lattice results for the $SU(3)$ 
flavor components of the lowest three $\Lambda(1/2^-)$ states can be reproduced in this model: the lowest one is consistent with a 
$\lambda$-mode excitation, as is expected from Eq.\,(\ref{eq:ex.energy}), while the next two are $\rho$ modes with the diquark spin 
fixed to 0 and 1, respectively. The lowest few negative-parity $\Lambda_c$ states are hence most naturally understood to have a quark-model-type 
structure.

\section*{Acknowledgments}
All the numerical calculations were performed on SR16000 at YITP, Kyoto
University and on TSUBAME at Tokyo Institute of Technology. The unquenched gauge configurations were all generated by 
PACS-CS Collaboration~\cite{Aoki:2008sm}. This work was supported in part by KAKENHI (Grants No. 25247036 and No. 16K05365). 


\appendix
\section{\label{NumRes} Numerical Results} 
In this appendix, we have collected the numerical results of this work. 

\subsection{Hadron masses}
Here, we provide the obtained hadron masses, together with 
their quadratically and linearly extrapolated values at the 
physical point. In Table~\ref{tab:num.eff.masses.1}, hadron masses for states containing 
only $u$ and $d$ quarks are given. Table~\ref{tab:num.eff.masses.2} lists the masses of 
kaons, $D$ mesons, as well as $\Sigma$ and $\Sigma_c$ baryons, together 
with corresponding states that have quark masses interpolating 
between strange and charm. Finally, Table~\ref{tab:num.eff.masses.3} gives the lowest two $\Lambda$ 
baryon masses with positive parity and the lowest four with negative 
parity, again with heavy quark masses ranging from strange to charm. 

\begin{widetext}
\begin{table}[h]
\renewcommand{\arraystretch}{1.3}
\begin{center}
\caption{Hadron masses for states containing no strange or charm valence quark ($\pi$ and $N$). 
Note that $\kappa_{s}$ is the hopping parameter of the strange sea quarks, and 
$\kappa_{ud}$ corresponds to the $u$ and $d$ quarks and approaches the 
physical value from top to bottom. 
The line denoted as ``phys. pt. (quad.)" gives the 
chirally extrapolated physical point results using a quadratic fit, while  
``phys. pt. (lin.)" gives the corresponding linear fit result. 
All values are given in lattice units.} 
\label{tab:num.eff.masses.1}
\begin{tabular}{cccccc}  
\hline 
$\kappa_{s}$& $\kappa_{ud}$ & $m_{\pi}$ & $m_N$ &  \\ \hline 
\hspace*{0.18cm}$0.13640$\hspace*{0.18cm} & \hspace*{0.18cm}$0.13700$ \hspace*{0.18cm}& 
\hspace*{0.18cm}$0.3220(11)$\hspace*{0.18cm} &
\hspace*{0.18cm} $0.715(11)$ \hspace*{0.18cm} \\[-4pt] 
                  & $0.13727$ & $0.2635(11)$ & $0.649(18)$ \\[-4pt]
                  & $0.13754$ & $0.1895(12)$ & $0.560(12)$  \\[-4pt] 
                  & $0.13770$ & $0.1323(11)$ & $0.518(17)$  \\[-4pt] 
                  & phys. pt. (quad.) & $0.0840(19)$ & $0.470(30)$ \\[-4pt]
                  & phys. pt. (lin.) & $0.1113(11)$ & $0.488(14)$ \\
\hline
\end{tabular}
\end{center}
\end{table}

\begin{table}
\renewcommand{\arraystretch}{1.3}
\begin{center}
\caption{$K$($D$) and $\Sigma_{(c)}$ 
hadron masses for different hopping parameter combinations, 
extrapolated to the physical point. Here, $\kappa_{sc}$ is the hopping parameter 
of the heavy quark, which is changed from strange (top) to charm (bottom). 
Note that $\kappa_{ud}$ corresponds to the $u$ and $d$ quark and approaches the 
physical value from top to bottom. 
The line denoted as ``phys. pt. (quad.)" in each block gives the 
chirally extrapolated physical point results using a quadratic fit, while  
``phys. pt. (lin.)" gives the corresponding linear fit result. 
All values are given in lattice units. 
} 
\label{tab:num.eff.masses.2}
\begin{tabular}{cccc}  
\hline 
$\kappa_{sc}$& $\kappa_{ud}$ & $m_{K (D)}$ & $m_{\Sigma_{(c)}}$  \\ \hline
\hspace*{0.80cm}$0.13640$\hspace*{0.80cm} & \hspace*{0.80cm}$0.13700$ \hspace*{0.80cm}& 
\hspace*{0.80cm}$0.3622(11)$\hspace*{0.80cm} & \hspace*{0.80cm}$0.751(10)$ \hspace*{0.80cm} \\[-4pt]
                  & $0.13727$ & $0.3300(11)$ & $0.705(16)$  \\[-4pt]
                  & $0.13754$ & $0.2948(10)$ & $0.637(12)$  \\[-4pt] 
                  & $0.13770$ & $0.2747(13)$ & $0.619(12)$ \\[-4pt] 
                  & phys. pt. (quad.) & $0.2563(22)$ & $0.591(23)$ \\[-4pt]
                  & phys. pt. (lin.) & $0.2618(11)$ & $0.592(11)$ \\[-4pt]
$0.13300$& $0.13700$ & $0.5435(16)$ & $0.914(11)$ \\[-4pt]
                 & $0.13727$ & $0.5210(15)$ & $0.879(13)$ \\[-4pt]           
                 & $0.13754$ & $0.4966(16)$ & $0.800(14)$ \\[-4pt]
                 & $0.13770$ & $0.4815(21)$ & $0.802(12)$ \\[-4pt]      
                 & phys. pt. (quad.) & $0.4685(33)$ & $0.773(21)$ \\[-4pt]  
                 & phys. pt. (lin.) & $0.4732(18)$ & $0.774(12)$ \\[-4pt]            
$0.12900$& $0.13700$ & $0.7103(20)$ & $1.068(13)$ \\[-4pt]
                 & $0.13727$ & $0.6913(18)$ & $1.043(14)$ \\[-4pt]           
                 & $0.13754$ & $0.6713(23)$ & $0.960(16)$ \\[-4pt]
                 & $0.13770$ & $0.6560(29)$ & $0.963(13)$ \\[-4pt]      
                 & phys. pt. (quad.) & $0.6450(44)$ & $0.929(23)$ \\[-4pt]
                 & phys. pt. (lin.) & $0.6505(24)$ & $0.938(13)$ \\[-4pt]   
$0.12600$& $0.13700$ & $0.8186(22)$ & $1.169(13)$ \\[-4pt]
                 & $0.13727$ & $0.8018(22)$ & $1.144(16)$ \\[-4pt]           
                 & $0.13754$ & $0.7841(24)$ & $1.078(12)$ \\[-4pt]
                 & $0.13770$ & $0.7663(35)$ & $1.068(14)$ \\[-4pt]      
                 & phys. pt. (quad.) & $0.7558(53)$ & $1.036(25)$ \\[-4pt]
                 & phys. pt. (lin.) & $0.7639(27)$ & $1.045(13)$ \\[-4pt]
$0.12240$& $0.13700$ & $0.9365(26)$ & $1.280(14)$ \\[-4pt]
                 & $0.13727$ & $0.9183(26)$ & $1.276(13)$ \\[-4pt]           
                 & $0.13754$ & $0.9047(31)$ & $1.192(13)$ \\[-4pt]
                 & $0.13770$ & $0.8880(47)$ & $1.187(13)$ \\[-4pt]      
                 & phys. pt. (quad.) & $0.8823(68)$ & $1.138(23)$ \\[-4pt]
                 & phys. pt. (lin.) & $0.8855(35)$ & $1.166(13)$ \\
\hline
\end{tabular}
\end{center}
\end{table}

\begin{table*}
\renewcommand{\arraystretch}{1.3}
\begin{center}
\caption{Same as in Table \ref{tab:num.eff.masses.2}, but for $\Lambda$ baryon 
states with spin $1/2$. Note that $E_i(1/2^{\pm})$ stands for the $i$th state of spin $1/2$ with parity $\pm$.} 
\label{tab:num.eff.masses.3}
\begin{tabular}{cccccccc}  
\hline 
$\kappa_{sc}$& $\kappa_{ud}$ & $E_{1}(1/2^{+})$ & $E_{2}(1/2^{+})$ 
& $E_{1}(1/2^{-})$ & $E_{2}(1/2^{-})$ & $E_{3}(1/2^{-})$ & $E_{4}(1/2^{-})$ \\ \hline
\hspace*{0.30cm}$0.13640$\hspace*{0.30cm} & \hspace*{0.30cm}$0.13700$ \hspace*{0.30cm}& 
\hspace*{0.30cm}$0.762(7)$\hspace*{0.30cm} & \hspace*{0.30cm}$1.321(25)$\hspace*{0.30cm} &
\hspace*{0.30cm} $1.027(14)$ \hspace*{0.30cm}& \hspace*{0.30cm}$1.090(17)$ \hspace*{0.30cm} & 
\hspace*{0.30cm}$1.131(19)$ \hspace*{0.30cm} & \hspace*{0.30cm}$1.527(47)$ \hspace*{0.30cm} \\[-4pt]
                  & $0.13727$ & $0.695(8)$ & $1.236(50)$ & $0.937(16)$ & $1.014(23)$ & $1.029(26)$ & $1.509(37)$\\[-4pt] 
                  & $0.13754$ & $0.643(8)$ & $1.318(59)$ & $0.811(33)$ & $0.947(25)$ & $1.023(23)$ & $1.486(42)$\\[-4pt] 
                  & $0.13770$ & $0.593(8)$ & $1.235(24)$ & $0.755(30)$ & $0.896(27)$ & $0.958(45)$ & $1.504(40)$\\[-4pt] 
                  & phys. pt. (quad.) & $0.564(14)$ & $1.242(54)$ & $0.685(45)$ & $0.862(45)$ & $1.000(64)$ & $1.504(69)$\\[-4pt] 
                  & phys. pt. (lin.) & $0.573(7)$ & $1.226(27)$ & $0.723(26)$ & $0.871(24)$ & $0.949(28)$ & $1.488(39)$\\[-4pt]                  
$0.13300$ & $0.13700$ & $0.898(9)$ & $1.447(24)$ & $1.165(12)$ & $1.217(16)$ & $1.252(19)$ & $1.635(47)$ \\[-4pt] 
                  & $0.13727$ & $0.837(8)$ & $1.400(19)$ & $1.012(31)$ & $1.149(20)$ & $1.168(26)$ & $1.654(50)$ \\[-4pt] 
                  & $0.13754$ & $0.795(7)$ & $1.346(18)$ & $0.965(31)$ & $1.099(32)$ & $1.163(17)$ & $1.591(41)$ \\[-4pt] 
                  & $0.13770$ & $0.737(10)$ & $1.262(39)$ & $0.884(32)$ & $1.060(35)$ & $1.089(25)$ & $1.598(38)$ \\[-4pt] 
                  & phys. pt. (quad.) & $0.713(16)$ & $1.235(52)$ & $0.874(55)$ & $1.040(52)$ & $1.083(44)$ & $1.568(72)$ \\[-4pt]  
                  & phys. pt. (lin.) & $0.729(8)$ & $1.281(24)$ & $0.840(28)$ & $1.036(29)$ & $1.093(20)$ & $1.587(38)$ \\[-4pt]   
$0.12900$ & $0.13700$ & $1.041(9)$ & $1.573(25)$ & $1.300(14)$ & $1.358(16)$ & $1.392(20)$ & $1.753(48)$ \\[-4pt] 
                  & $0.13727$ & $0.981(9)$ & $1.526(19)$ & $1.149(32)$ & $1.293(21)$ & $1.327(28)$ & $1.800(51)$\\[-4pt] 
                  & $0.13754$ & $0.947(8)$ & $1.477(18)$ & $1.112(32)$ & $1.230(32)$ & $1.308(14)$ & $1.714(41)$ \\[-4pt] 
                  & $0.13770$ & $0.883(11)$ & $1.396(43)$ & $1.070(23)$ & $1.202(32)$ & $1.240(26)$ & $1.711(37)$ \\[-4pt] 
                  & phys. pt. (quad.) & $0.864(18)$ & $1.377(55)$ & $1.074(44)$ & $1.175(50)$ & $1.220(45)$ & $1.659(71)$ \\[-4pt]
                  & phys. pt. (lin.) & $0.880(9)$ & $1.416(25)$ & $1.021(23)$ & $1.173(28)$ & $1.250(18)$ & $1.704(37)$ \\[-4pt]                
$0.12600$ & $0.13700$ & $1.138(10)$ & $1.660(27)$ & $1.392(15)$ & $1.453(16)$ & $1.489(21)$ & $1.837(48)$ \\[-4pt] 
                  & $0.13727$ & $1.080(9)$ & $1.611(21)$ & $1.248(32)$ & $1.375(31)$ & $1.383(21)$ & $1.870(51)$ \\[-4pt] 
                  & $0.13754$ & $1.043(9)$ & $1.563(18)$ & $1.195(35)$ & $1.322(33)$ & $1.400(14)$ & $1.798(41)$ \\[-4pt] 
                  & $0.13770$ & $0.981(12)$ & $1.488(43)$ & $1.165(24)$ & $1.301(34)$ & $1.337(26)$ & $1.794(37)$ \\[-4pt] 
                  & phys. pt. (quad.) & $0.960(20)$ & $1.468(57)$ & $1.167(45)$ & $1.288(57)$ & $1.373(43)$ & $1.750(71)$ \\[-4pt]
                  & phys. pt. (lin.) & $0.976(10)$ & $1.503(26)$ & $1.114(24)$ & $1.266(30)$ & $1.341(18)$ & $1.786(37)$ \\[-4pt]      
$0.12240$ & $0.13700$ & $1.245(10)$ & $1.755(30)$ & $1.496(16)$ & $1.559(16)$ & $1.596(21)$ & $1.934(49)$ \\[-4pt] 
                  & $0.13727$ & $1.188(10)$ & $1.720(24)$ & $1.438(17)$ & $1.482(21)$ & $1.483(33)$ & $1.970(52)$ \\[-4pt] 
                  & $0.13754$ & $1.152(10)$ & $1.664(18)$ & $1.364(19)$ & $1.425(34)$ & $1.506(14)$ & $1.897(42)$\\[-4pt] 
                  & $0.13770$ & $1.088(13)$ & $1.617(27)$ & $1.270(26)$ & $1.411(35)$ & $1.439(26)$ & $1.891(37)$\\[-4pt] 
                  & phys. pt. (quad.) & $1.066(21)$ & $1.577(46)$ & $1.215(39)$ & $1.397(53)$& $1.446(49)$ & $1.845(72)$\\[-4pt]     
                  & phys. pt. (lin.) & $1.086(11)$ & $1.609(23)$ & $1.273(20)$ & $1.370(29)$& $1.446(26)$ & $1.884(37)$\\
\hline
\end{tabular}
\end{center}
\end{table*}
\clearpage

\subsection{Couplings}
In this subsection, the normalized $SU(3)$ coupling strengths defined 
in Eqs.~(\ref{eq:singletoctet1}) and (\ref{eq:singletoctet2}) are listed for all $\Lambda$ baryon states, which 
could be extracted with a sufficiantly clear signal. In Table~\ref{tab:num.coupl.1}, the 
couplings for the lowest $\Lambda$ baryon state with positive parity are 
given. The table contains the couplings for the physical $\Lambda$ and 
$\Lambda_c$ states as well as for states with unphysical quarks that 
interpolate between strange and charm. Table~\ref{tab:num.coupl.2}, is the same as Table~\ref{tab:num.coupl.1}, 
but for the lowest three $\Lambda$ baryon states with negative parity.
\begin{table}[h]
\renewcommand{\arraystretch}{1.3}
\begin{center}
\caption{Normalized coupling strength of $\Lambda$ baryons with spin $1/2$ and 
positive parity to singlet and octet operators. Note that 
$g_i^{\,\bm{1}/\bm{8}}(1/2^{\pm})$ stands for the normalized coupling of the $i$-th state of spin $1/2$ with parity $\pm$ to singlet ($\bm{1}$) or octet ($\bm{8}$) 
operators, as defined in Eqs.~(\ref{eq:singletoctet1}) and (\ref{eq:singletoctet2}). 
The line denoted as ``phys. pt. (quad.)" in each block gives the 
chirally extrapolated physical point results using a quadratic fit, while  
``phys. pt. (lin.)" gives the corresponding linear fit result.} 
\label{tab:num.coupl.1}
\begin{tabular}{cccc}  
\hline 
$\kappa_{sc}$& $\kappa_{ud}$ & $g_{1}^{\,\bm{1}}(1/2^{+})$ & $g_{1}^{\,\bm{8}}(1/2^{+})$  \\ \hline
\hspace*{0.80cm}$0.13640$\hspace*{0.80cm} & \hspace*{0.80cm}$0.13700$ \hspace*{0.80cm}& 
\hspace*{0.80cm}$0.000(0)$\hspace*{0.80cm} & \hspace*{0.80cm}$1.000(63)$\hspace*{0.80cm} \\[-4pt] 
                  & $0.13727$ & $0.002(1)$ & $0.998(136)$ \\[-4pt] 
                  & $0.13754$ & $0.003(2)$ & $0.997(81)$ \\[-4pt] 
                  & $0.13770$ & $0.002(2)$ & $0.998(108)$ \\ [-4pt]
                  & phys. pt. (quad.) & $0.002(3)$ & $0.998(199)$ \\[-4pt]
                  & phys. pt. (lin.) & $0.004(1)$ & $0.996(88)$ \\[-4pt] 
$0.13300$ & $0.13700$ & $0.003(2)$ & $0.997(32)$ \\[-4pt] 
                  & $0.13727$ & $0.001(1)$ & $0.999(10)$ \\[-4pt] 
                  & $0.13754$ & $0.004(2)$ & $0.996(21)$ \\[-4pt] 
                  & $0.13770$ & $0.003(2)$ & $0.997(14)$ \\[-4pt] 
                  & phys. pt. (quad.) & $0.005(3)$ & $0.995(25)$ \\[-4pt]
                  & phys. pt. (lin.) & $0.003(2)$ & $0.997(16)$ \\[-4pt]               
$0.12900$ & $0.13700$ & $0.006(6)$ & $0.994(52)$ \\[-4pt] 
                  & $0.13727$ & $0.004(4)$ & $0.996(17)$ \\[-4pt] 
                  & $0.13754$ & $0.011(7)$ & $0.989(38)$ \\[-4pt] 
                  & $0.13770$ & $0.008(5)$ & $0.992(24)$ \\[-4pt] 
                  & phys. pt. (quad.) & $0.011(9)$ & $0.989(42)$ \\[-4pt]
                  & phys. pt. (lin.) & $0.009(6)$ & $0.991(27)$ \\[-4pt]                  
$0.12600$ & $0.13700$ & $0.010(11)$ & $0.990(65)$ \\[-4pt] 
                  & $0.13727$ & $0.006(7)$ & $0.994(26)$ \\[-4pt] 
                  & $0.13754$ & $0.019(12)$ & $0.981(50)$ \\[-4pt]
                  & $0.13770$ & $0.011(9)$ & $0.989(32)$ \\[-4pt] 
                  & phys. pt. (quad.) & $0.017(15)$ & $0.983(57)$ \\[-4pt]
                  & phys. pt. (lin.) & $0.013(9)$ & $0.987(36)$ \\[-4pt]            
$0.12240$ & $0.13700$ & $0.015(17)$ & $0.985(81)$ \\[-4pt] 
                  & $0.13727$ & $0.007(11)$ & $0.993(34)$ \\[-4pt] 
                  & $0.13754$ & $0.027(19)$ & $0.973(65)$ \\[-4pt] 
                  & $0.13770$ & $0.014(12)$ & $0.986(40)$ \\[-4pt] 
                  & phys. pt. (quad.) & $0.021(21)$ & $0.979(71)$ \\[-4pt]
                  & phys. pt. (lin.) & $0.017(13)$ & $0.983(45)$ \\       
\hline
\end{tabular}
\end{center}
\end{table}

\begin{table*}
\renewcommand{\arraystretch}{1.3}
\begin{center}
\caption{Same as in Table \ref{tab:num.coupl.1}, but for $\Lambda$ baryons with negative parity.} 
\label{tab:num.coupl.2}
\begin{tabular}{cccccccc}  
\hline 
$\kappa_{sc}$& $\kappa_{ud}$ 
& $g_{1}^{\,\bm{1}}(1/2^{-})$ & $g_{1}^{\,\bm{8}}(1/2^{-})$ & $g_{2}^{\,\bm{1}}(1/2^{-})$ & $g_{2}^{\,\bm{8}}(1/2^{-})$ 
& $g_{3}^{\,\bm{1}}(1/2^{-})$ & $g_{3}^{\,\bm{8}}(1/2^{-})$ \\ \hline
\hspace*{0.30cm}$0.13640$\hspace*{0.30cm} & \hspace*{0.30cm}$0.13700$ \hspace*{0.30cm}& 
\hspace*{0.30cm}$0.984(3)$\hspace*{0.30cm} & \hspace*{0.30cm}$0.016(2)$\hspace*{0.30cm} &
\hspace*{0.30cm} $0.005(2)$ \hspace*{0.30cm}& \hspace*{0.30cm}$0.995(212)$ \hspace*{0.30cm} & 
\hspace*{0.30cm}$0.001(1)$ \hspace*{0.30cm} & \hspace*{0.30cm}$0.999(127)$ \hspace*{0.30cm} \\[-4pt] 
                  & $0.13727$ & $0.963(8)$ & $0.037(7)$ & $0.007(3)$ & $0.993(235)$ & $0.001(2)$ & $0.999(358)$ \\[-4pt] 
                  & $0.13754$ & $0.941(13)$ & $0.059(10)$ & $0.013(7)$ & $0.987(209)$ & $0.014(6)$ & $0.986(129)$ \\[-4pt] 
                  & $0.13770$ & $0.944(15)$ & $0.056(11)$ & $0.013(6)$ & $0.987(125)$ & $0.011(3)$ & $0.989(93)$ \\[-4pt] 
                  & phys. pt. (quad.) & $0.936(22)$ & $0.064(17)$ & $0.017(9)$ & $0.983(269)$ & $0.017(5)$ & $0.983(231)$ \\[-4pt]  
                  & phys. pt. (lin.) & $0.929(11)$ & $0.071(9)$ & $0.014(5)$ & $0.986(140)$ & $0.010(3)$ & $0.990(99)$ \\[-4pt]  
$0.13300$ & $0.13700$ & $0.805(47)$ & $0.195(38)$ & $0.032(13)$ & $0.968(123)$ & $0.011(4)$ & $0.989(81)$ \\[-4pt]  
                  & $0.13727$ & $0.786(44)$ & $0.214(36)$ & $0.044(15)$ & $0.956(67)$ & $0.019(55)$ & $0.981(125)$ \\[-4pt]   
                  & $0.13754$ & $0.744(51)$ & $0.256(43)$ & $0.044(26)$ & $0.956(167)$ & $0.037(10)$ & $0.963(77)$ \\[-4pt]  
                  & $0.13770$ & $0.769(49)$ & $0.231(37)$ & $0.043(35)$ & $0.957(55)$ & $0.057(13)$ & $0.943(105)$ \\[-4pt]  
                  & phys. pt. (quad.) & $0.763(83)$ & $0.237(64)$ & $0.037(48)$ & $0.963(107)$ & $0.071(18)$ & $0.929(192)$ \\[-4pt]
                  & phys. pt. (lin.) & $0.748(47)$ & $0.252(36)$ & $0.051(25)$ & $0.949(65)$ & $0.052(9)$ & $0.948(88)$ \\[-4pt]               
$0.12900$ & $0.13700$ & $0.656(55)$ & $0.344(50)$ & $0.046(18)$ & $0.954(40)$ & $0.024(10)$ & $0.976(92)$ \\[-4pt]  
                  & $0.13727$ & $0.665(51)$ & $0.335(45)$ & $0.075(20)$ & $0.925(76)$ & $0.020(13)$ & $0.980(231)$ \\[-4pt]  
                  & $0.13754$ & $0.660(52)$ & $0.340(46)$ & $0.066(29)$ & $0.934(62)$ & $0.067(18)$ & $0.933(111)$ \\[-4pt]  
                  & $0.13770$ & $0.663(48)$ & $0.337(38)$ & $0.075(42)$ & $0.925(81)$ & $0.095(18)$ & $0.905(143)$ \\[-4pt]  
                  & phys. pt. (quad.) & $0.659(86)$ & $0.341(69)$ & $0.055(59)$ & $0.945(137)$ & $0.136(29)$ & $0.864(281)$ \\[-4pt]
                  & phys. pt. (lin.) & $0.664(47)$ & $0.336(39)$ & $0.085(30)$ & $0.915(65)$ & $0.090(16)$ & $0.910(120)$ \\[-4pt]  
$0.12600$ & $0.13700$ & $0.590(52)$ & $0.410(50)$ & $0.048(19)$ & $0.952(53)$ & $0.034(16)$ & $0.966(106)$ \\[-4pt]  
                  & $0.13727$ & $0.624(13)$ & $0.376(36)$ & $0.092(21)$ & $0.908(68)$ & $0.070(14)$ & $0.930(145)$ \\[-4pt]  
                  & $0.13754$ & $0.610(52)$ & $0.390(47)$ & $0.058(17)$ & $0.942(25)$ & $0.075(19)$ & $0.925(56)$ \\[-4pt]  
                  & $0.13770$ & $0.614(45)$ & $0.386(36)$ & $0.060(20)$ & $0.940(30)$ & $0.122(21)$ & $0.878(144)$ \\[-4pt]  
                  & phys. pt. (quad.) & $0.599(81)$ & $0.401(69)$ & $0.028(36)$ & $0.972(66)$ & $0.127(34)$ & $0.873(250)$ \\[-4pt]
                  & phys. pt. (lin.) & $0.619(45)$ & $0.381(38)$ & $0.064(18)$ & $0.936(30)$ & $0.119(19)$ & $0.881(84)$ \\[-4pt]      
$0.12240$ & $0.13700$ & $0.548(50)$ & $0.452(47)$ & $0.101(30)$ & $0.899(61)$ & $0.047(22)$ & $0.953(122)$ \\[-4pt]  
                  & $0.13727$ & $0.586(49)$ & $0.414(50)$ & $0.096(24)$ & $0.904(81)$ & $0.082(15)$ & $0.918(156)$ \\[-4pt]  
                  & $0.13754$ & $0.573(49)$ & $0.427(44)$ & $0.080(30)$ & $0.920(75)$ & $0.110(26)$ & $0.890(103)$ \\[-4pt]  
                  & $0.13770$ & $0.577(39)$ & $0.423(34)$ & $0.058(30)$ & $0.942(82)$ & $0.154(22)$ & $0.846(68)$ \\[-4pt]  
                  & phys. pt. (quad.) & $0.562(74)$ & $0.438(66)$ & $0.041(50)$ & $0.959(142)$ & $0.178(36)$ & $0.822(156)$ \\[-4pt]
                  & phys. pt. (lin.) & $0.584(41)$ & $0.416(35)$ & $0.059(28)$ & $0.941(74)$ & $0.162(22)$ & $0.838(76)$ \\                 
\hline
\end{tabular}
\end{center}
\end{table*}
\end{widetext}

\end{document}